# A deep-learning model for predicting daily $PM_{2.5}$ concentration in response to emission reduction

Shigan Liu[1], Guannan Geng[2], Yanfei Xiang[1], Hejun Hu[1], Xiaodong Liu[2], Xiaomeng Huang[1], and Qiang Zhang[1]

[1] Ministry of Education Key Laboratory for Earth System Modeling, Department of Earth System Science, Tsinghua University, Beijing 100084, China

[2] State Key Laboratory of Regional Environment and Sustainability, School of Environment, Tsinghua University, Beijing 100084, China

**Correspondence to: Guannan Geng** (guannangeng@tsinghua.edu.cn)**, Xiaomeng Huang** (hxm@tsinghua.edu.cn)**, or Qiang Zhang** (qiangzhang@tsinghua.edu.cn)

**Main Text:**

12 pages of text (excluding references, and figure legends)

Figs. 1–5

**Supplementary Online Materials:**

Methods

Supplementary Figures 1–9

Supplementary Tables 1–5

**Abstract**

Air pollution remains a leading global health threat, with fine particulate matter ($PM_{2.5}$) contributing to millions of premature deaths annually. Chemical transport models (CTMs) are essential tools for evaluating how emission controls improve air quality and save lives, but they are computationally intensive. Reduced form models accelerate simulations but sacrifice spatial-temporal granularity, accuracy, and flexibility. Here we present CleanAir, a deep-learning-based model developed as an efficient alternative to CTMs in simulating daily $PM_{2.5}$ and its chemical compositions in response to precursor emission reductions at 36 km resolution, which could predict $PM_{2.5}$ concentration for a full year within 10 seconds on a single GPU, a speed five orders of magnitude faster. Built on a Residual Symmetric 3D U-Net architecture and trained on more than 2,400 emission reduction scenarios generated by a well-validated Community Multiscale Air Quality (CMAQ) model, CleanAir generalizes well across unseen meteorological years and emission patterns. It produces results comparable to CMAQ in both absolute concentrations and emission-induced changes, enabling efficient, full-coverage simulations across short-term interventions and long-term planning horizons. This advance empowers researchers and policymakers to rapidly evaluate a wide range of air quality strategies and assess the associated health impacts, thereby supporting more responsive and informed environmental decision-making.

**Main text**

Air pollution poses a grave threat to human health and ecosystems[1–3]: fine particulate matter ($PM_{2.5}$) alone contributes to millions of premature deaths each year and exacerbates cardiovascular and respiratory diseases worldwide[4]. Chemical transport models (CTMs) are essential tools to predict how emission controls translate into cleaner air and lives saved[5–7]. By solving coupled chemical (e.g. production and loss) and physical (e.g., advection and deposition) processes in the atmosphere through partial differential equations (PDEs), CTMs can quantify the nonlinear responses of air pollutant concentrations to precursor emission changes[8,9]. Yet this simulation comes at a cost[10]: simulating a single full year air quality over China at 36 km resolution can demand days of supercomputer time, severely limiting the speed and scope of scenario analysis.

To reduce computational burden, reduced-form models (RFMs), whether statistical surrogates or machine-learning emulators, have been developed to approximate the complex, nonlinear relationships between emissions and concentrations[11–20]. Although RFMs accelerate simulations, they usually sacrifice CTM versatility and flexibility. For example, RFMs are typically calibrated to specific meteorological years within the training dataset and perform poorly outside those conditions[11–14,16,18,20]. To reduce modeling complexity, RFMs have to trade off temporal resolution, spatial resolution, and model accuracy[11–20]. Models that allow for spatial heterogeneity in emission changes output only annual mean concentrations[11–14,18,20], whereas those that preserve daily resolution assume uniform emission changes across an entire region[15,19]. Moreover, the use of simplifications (e.g., linear, polynomial) when modeling emission-concentration relationships further undermines predictive accuracy[11,12,19,20].

Recent breakthroughs in artificial intelligence (AI) for weather forecasting[21–25] have demonstrated that deep neural networks can learn complex, high-dimensional dynamics at unprecedented speed. AI techniques have been applied to air quality forecasting, the Aurora[26], which produces short-term forecasts of aerosol and reactive gas concentrations several days in advance. However, Aurora's outputs still exhibit notable biases compared with reanalysis datasets, and it cannot be considered a true air-quality model because it lacks support for continuous, long-term simulations (for example, a full year)[26]. To date, no AI-based model matches CTMs in spatiotemporal resolution, multi-year simulation capacity, and the flexibility to handle diverse emission scenarios.

Here we present CleanAir, a deep-learning model that bridges this gap by enabling continuous, daily-scale simulation of $PM_{2.5}$ and its chemical components over multi-year periods, at a speed over 40,000 times faster—a major step forward for AI in air quality modeling. Developed as an efficient alternative to physics-based CTMs, CleanAir works at 36 km resolution to predict $PM_{2.5}$ and its chemical compositions in response to precursor emission reductions (domain shown in Supplementary Figure 1). We focus on $PM_{2.5}$ as it remains the most pressing air pollution challenge in China, with complex, nonlinear formation pathways involving multiple pollutants and sources. CleanAir is built on a Residual Symmetric 3D U-Net architecture[27] and trained on a large-scale daily dataset of hypothetical emission reduction scenarios generated by the Community Multiscale Air Quality (CMAQ) model (see more details in the *Methods* section). Leveraging the powerful representational capacity of deep learning, the model achieves strong generalization across unseen meteorological conditions and emission patterns, extending its applicability beyond the training conditions. Our evaluations confirm that CleanAir reliably reproduces both absolute concentrations and emission-induced changes in $PM_{2.5}$ across temporal and spatial scales, closely matching CMAQ's performance. Yet it requires only 9 seconds to complete an annual simulation, representing a five-order-of-magnitude speedup. This dramatic improvement in computational efficiency makes CleanAir particularly valuable for policy assessment, multi-scenario optimization, and time-sensitive applications such as short-term emission control.

**The CleanAir model for $PM_{2.5}$ predictions**

Figure 1a–c presents a schematic of the CleanAir model. The model takes five key inputs (see details in *Methods* and Supplementary Figure 2): (1) anthropogenic emissions fixed at 2017 levels from the Multiresolution Emission Inventory for China – High Resolution (MEIC-HR, http://meicmodel.org.cn)[28], used as the baseline; (2) biogenic emissions at any given time of interest, generated by MEGAN[29]; (3) meteorological fields simulated by the Weather Research and Forecasting (WRF, https://www.mmm.ucar.edu/models/wrf) at the given time; (4) baseline simulations of $PM_{2.5}$ components and other relevant species (e.g., reactive intermediates listed in Supplementary Table 1), produced by CMAQ using 2017 anthropogenic emissions along with biogenic emissions and meteorological data at the given time; and (5) a new anthropogenic emission scenario to be simulated that are comparable to or lower than the baseline. The input emission is categorized into three emission layers, i.e., surface (transportation, residential and agriculture), intermediate (industry) and elevated (power plants with elevated stacks) sources, due to their differing impacts on surface-level concentrations. It includes eight air pollutant species closely linked to $PM_{2.5}$ formation: sulfur dioxide ($SO_2$), nitrogen oxides ($NO_x$), primary fine particulate matter ($PM_{2.5}$), black carbon (BC), organic carbon (OC), coarse particulate matter with aerodynamic diameter between 2.5 and 10 μm ($PM_{coarse}$), ammonia ($NH_3$), and non-methane volatile organic compounds (NMVOCs). The model outputs $PM_{2.5}$ chemical component concentrations, including sulfate ($SO_4^{2-}$), nitrate ($NO_3^-$), ammonium ($NH_4^+$), organic matter (OM), black carbon (BC), and other components ($PM_{other}$) in surface layer at daily scale,

under the specified meteorological conditions and the given emission scenario. It is worth noting that, inputs (1)–(4) have been pre-simulated using CMAQ based on 2017 MEIC-HR emissions with biogenic emissions and meteorological data varying from 2000 onward. These pre-simulated datasets are embedded in the model as default inputs, allowing the model to support simulations under meteorological conditions from 2000 to the present without additional CMAQ runs during model application.

The model is built upon the architecture of the Residual Symmetric 3D U-Net[27] (Fig 1a and 1b) and is trained using a large dataset generated by CMAQ that captures the relationship between $PM_{2.5}$ concentration changes and precursor emission reductions (Fig. 1c and 1d). The chemical composition of $PM_{2.5}$ is complex, and its concentration is influenced by the combined effects of multiple precursor emissions involving nonlinear atmospheric chemistry[30], making it especially important to construct a representative and comprehensive training dataset. Our dataset was created by perturbing the 2017 baseline emissions across 15 dimensions, that is, five pollutant species (assuming primary PM components such as $PM_{2.5}$, BC, OC, and $PM_{coarse}$ share the same reduction rate to reduce modeling complexity) across three emission layers, by reducing each from 0% to 100% (Fig 1d). In total, 2,416 emission reduction scenarios were created (Supplementary Figure 3). Of these, 1,728 were generated through random perturbations across the 15 dimensions, with Sobol's algorithm[31] applied to ensure a uniform distribution of variations. Additional 688 scenarios were added to specifically target (1) individual emission layers (to account for vertical sensitivity differences), and (2) $NO_x$ and NMVOCs emissions (to capture the nonlinear chemical responses of secondary aerosol formation). These emission scenarios were then simulated using CMAQ for one month in January, April, July or October of 2017, representing winter, spring, summer, and autumn, respectively. This seasonal sampling accounts for the variability of meteorological influences throughout the year, enhancing the model's ability to generalize across diverse atmospheric conditions. The above dataset was randomly split into 60% for training, 20% for validation, and 20% for testing.

The model architecture begins by feeding the input data into head layers for shape alignment and preliminary feature extraction. These inputs, which include data from CMAQ's different grid configurations, are processed through two parallel branches of 3D convolutional layers (Conv3D) with a padding layer (Padding) to unify dimensions. The extracted features are then passed through a symmetric U-Net structure comprising five residual modules with max-pooling and up-sampling layers, as well as skip connections using summation joins to facilitate feature fusion. Each residual module consists of three convolutional blocks of Conv3D and Group Normalization (GroupNorm), each followed by an Exponential Linear Unit (ELU) activation. Skip connections within each module form residual subnetworks that enhance gradient flow and improve training efficiency. After advanced feature processing, the outputs are passed through tail layers, comprising a Conv3D layer and multiple 2D convolutional layers (Conv2D) with Rectified Linear Unit (ReLU) activation, to generate predicted changes in pollutant concentrations relative to the baseline scenario. The final concentrations for the emission reduction scenario are obtained by adding these predicted changes to the baseline concentrations. It is worth noting that while some other reactants are included in the model outputs, they are used solely as chemical constraints during training and do not serve as predictive targets.

The CleanAir model was trained using CMAQ simulation outputs as supervisory signals (Fig. 1c). During each training epoch, CleanAir was iteratively applied to the training dataset to generate predictions for $PM_{2.5}$ components and other reactants. These predictions were then compared to the corresponding CMAQ outputs using a loss function (see *Methods*; Supplementary Figure 4), and model parameters were updated via gradient backpropagation. The model was trained for 16 epochs, requiring approximately 40 hours in total. To mitigate the risk of overfitting, the training samples were randomly shuffled at the beginning of each epoch. Given that the model outputs span multiple pollutant components, a key challenge was ensuring balanced learning across all outputs. To address this, we implemented an adaptive weighted loss function, drawing on principles from multi-task learning[32,33]. Further details on the loss function design and training strategy are provided in the *Methods* section.

**Model performance**

We assessed CleanAir's performance in simulating $PM_{2.5}$ concentrations in response to emission reductions by comparing its outputs with those of CMAQ using the test dataset described above. Both the absolute concentrations of $PM_{2.5}$ and its chemical components, as well as their changes relative to the baseline (Δ), were evaluated at monthly and daily scales. Evaluation metrics included Pearson's correlation coefficient (R) and root-mean-squared error (RMSE), with results presented in Fig. 2 and Supplementary Figure 5–6.

Grid-level comparison of monthly-scale $\Delta PM_{2.5}$ concentrations (Fig. 2a) simulated by CleanAir and CMAQ shows excellent agreement, with an R of 0.999 and an RMSE of 0.281 μg m$^{-3}$. At the daily scale (Supplementary Figure 5a), the agreement remains strong, with an R of 0.998 and a higher RMSE of 0.582 μg m$^{-3}$, indicating that CleanAir effectively captures $PM_{2.5}$ concentration responses at fine temporal scales. Notably, CleanAir occasionally simulate positive $\Delta PM_{2.5}$ values reaching up to 10 μg m$^{-3}$ in response to emission reductions, consistent with CMAQ results. This suggests that CleanAir is capable of capturing the nonlinear chemical responses that can lead to counterintuitive outcomes under certain conditions. CleanAir also demonstrates strong consistency with CMAQ in simulating individual $PM_{2.5}$ chemical components, capturing their distinct concentration-emission response characteristics (Fig. 2b, Supplementary Figure 5b and Supplementary Figure 6). Primary components such as $\mathrm{BC}$ and $\mathrm{PM_{other}}$ exhibit slightly better performance than secondary components like $\mathrm{SO_4^{2-}}$, $\mathrm{NO_3^-}$, and $\mathrm{NH_4^+}$, likely due to the latter's more complex chemical formation pathways. CleanAir maintained high accuracy in clean conditions ($PM_{2.5}$ < 10 μg m$^{-3}$) and heavily polluted conditions ($PM_{2.5}$ > 115 μg m$^{-3}$), achieving R values above 0.995 and RMSE values below 2.02 μg m$^{-3}$ (Fig. 2c). These results indicate that CleanAir performs well across concentration extremes, comparable to CMAQ in capturing both low-end and high-end responses, which is particularly valuable for evaluating extreme pollution episodes and associated health risks.

We evaluate the inference speed of CleanAir on an NVIDIA RTX 5090 GPU (Gigabyte GV-N5090WF3OC-32GD, 32 GB) platform. Predicting one year of daily $PM_{2.5}$ and its chemical composition over China at a 36 km horizontal resolution only takes 9 seconds using a single GPU. In comparison, the CMAQ model requires approximately 4.5 days on a supercomputer equipped with 40 logical CPUs (2 × Intel(R) Xeon(R) Gold 6230 @ 2.10 GHz, 20 cores per socket) and 187 GB of physical random-access memory (RAM) for the same task, making CleanAir over 40,000 times faster.

**Modeling skills with different meteorological conditions and emission inventories**

Although the training dataset was constructed based on the four months of meteorological fields in 2017 and the MEIC-HR inventory, CleanAir has the capability to simulate $PM_{2.5}$ concentrations with meteorological conditions in different years and emission inventories with emission levels comparable to or lower than those of the 2017 MEIC-HR (see national emission totals from MEIC-HR in Supplementary Table 2). This flexibility represents a notable advancement over previous deep learning-based models, which were typically limited to the specific inventories or meteorological conditions[14,15,18,26].

To evaluate the flexibility on different meteorological and emission data, we use CleanAir to simulate $PM_{2.5}$ concentrations from 2017 to 2020 driven by year-by-year meteorological conditions and emission inventories (i.e., MEIC emission inevntory[7]). The results were compared against CMAQ simulations and ground-level observations from the China National Environmental Monitoring Centre (CNEMC; https://air.cnemc.cn:18007). While the 2017 MEIC and MEIC-HR inventories have similar national total emissions, their gridded spatial distributions differ: MEIC-HR incorporates location-specific emissions from over 100,000 industrial facilities, whereas MEIC allocates emissions using spatial proxies such as population density. From 2018 to 2020, MEIC emissions declined year over year due to China's ongoing clean air policies (Fig. 3a), and all are lower than the 2017 MEIC-HR baseline.

Fig. 3d shows that CleanAir demonstrates well agreement with ground observations at daily scale across multiple statistical metrics, with R over 0.6, RMSE below 35 μg $m^{-3}$ and normalized mean bias (NMB) within ±15%. These performance metrics are comparable to those of CMAQ and fall within the commonly recommended thresholds of NMB ≤ ±30% by Emery et al. (2016)[34] and NMB ≤ ±20% by Huang et al. (2021)[35], indicating that CleanAir reaches the current state-of-the-art standard for $PM_{2.5}$ modeling. Notably, model performance remains consistent between 2017 and other years, underscoring its robust generalization to unseen meteorological conditions. The discrepancy against observations is largely inherited from the CMAQ model itself, which serves as the training source. As such, it reflects the inherent limitations of the CTM rather than deficiencies in our deep learning model.

CleanAir also successfully captures both the spatial distribution (Fig. 3b) and interannual trends (Fig. 3c) of $PM_{2.5}$ concentrations across China at a 36 km spatial resolution, closely matching the CMAQ results. Spatially, the model reproduces key features such as the east–west gradient and major pollution hotspots such as the Beijing-Tianjin-Hebei area and the Sichuan Basin. Temporally, it reflects the year-to-year decline in $PM_{2.5}$ concentrations from 2017 to 2020, estimating a 19% reduction in population-weighted annual mean concentrations—closely aligned with CMAQ's estimate of 20%. These results affirm CleanAir's ability to generalize to new meteorological conditions and emission inventories while maintaining robust performance in daily $PM_{2.5}$ simulations.

**Model capability in short-term emission control**

Short-term emission controls are typically implemented during haze episodes, when unfavorable meteorological conditions lead to sharp increases in pollutant concentrations, and during major events to reduce air pollution in host cities and surrounding areas[36]. These temporary measures aim to reduce peak $PM_{2.5}$ levels and protect public health[30]. In such cases, rapid air quality simulations at daily scale are essential to support timely decision-making and ensure the effectiveness of interventions.

To demonstrate CleanAir's applicability in this context, we used both CleanAir and CMAQ to simulate a hypothetical short-term emission reduction case in February 2017 in China, during which two major pollution episodes occurred in the North China Plain. Three scenarios with increasing emission reduction intensity were designed, affecting 57 cities across six provinces (Beijing, Tianjin, Hebei, Shandong, Shanxi, and Henan; Fig. 4a, Supplementary Figure 1, and Supplementary Table 3). Targeted pollutants include $SO_2$, $NO_x$, primary $PM_{2.5}$, and NMVOCs, primarily from power, industry, residential, and transportation sectors. $NH_3$, mostly from agriculture sources, remains unchanged due to challenges in short-term regulation.

Fig. 4c shows that CleanAir accurately reproduces two major pollution episodes in February, with simulated peak $PM_{2.5}$ concentrations closely matching ground observations and CMAQ results. For instance, peak levels during the first and second episodes are estimated at 174 and 149 $\mu g\, m^{-3}$ by CleanAir, identical to CMAQ and within 1-6 $\mu g\, m^{-3}$ differences of observed values. This demonstrates that CleanAir is capable of capturing short-term pollution dynamics with accuracy comparable to a full CTM.

Building on this capability, CleanAir effectively predicts the air quality improvements under three emission control scenarios (Fig. 4b). In the aggressive control scenario, it estimates 224 additional city-level days below the national daily $PM_{2.5}$ standard (75 μg $m^{-3}$), compared to 254 days estimated by CMAQ—a 12% difference. Peak concentration reductions for the two episodes are also consistent between models, with CleanAir estimating decreases of up to 42 and 37 μg $m^{-3}$, respectively. Both models capture similar spatial mitigation patterns (Fig. 4d), with the most significant reductions concentrated over the Shandong, Henan, and Southeast of Hebei provinces.

Beyond accuracy, CleanAir offers substantial gains in computational efficiency. Simulating daily $PM_{2.5}$ concentrations under all three scenarios for one month requires only 2.2 seconds using one RTX 5090 GPU, whereas CMAQ needs approximately 27 hours. This speed advantage enables near real-time scenario evaluation, substantially lowering the cost of short-term policy assessment and supporting rapid decision-making for emergency air quality interventions.

**Model capability in long-term pollution intervention**

In air pollution mitigation studies, it is common to construct future emission pathways by designing policy scenarios with varying levels of implementation under different socioeconomic development trajectories. CTMs are then used to simulate the impact of these projected emissions on air pollutant concentrations, enabling the evaluation of policy effectiveness. Efficient and accurate simulation of $PM_{2.5}$ concentrations is therefore essential for exploring a wide range of policy combinations and their potential to improve air quality.

Here, we use emission scenarios derived from the Dynamic Projection model for Emissions in China version 1.2 (DPEC v1.2; http://meicmodel.org.cn)[37] to evaluate CleanAir's capability in simulating $PM_{2.5}$ concentrations under long-term emission trajectories. The five DPEC v1.2 scenarios represent varying combinations of air pollution control and carbon mitigation efforts by 2060, ranging from a reference scenario with no additional policies to progressively more ambitious pathways that integrate strengthened pollution controls, carbon peak targets, and carbon neutrality goals. Both CleanAir and CMAQ were used to simulate $PM_{2.5}$ concentrations, and the associated mortality risks were estimated using exposure-response functions from the Global Burden of Disease (GBD) 2019 study.

Fig. 5a and 5b present projections of China's $PM_{2.5}$ air quality and associated health impacts under five emission pathways simulated by CleanAir from 2020 to 2060, with CMAQ-based results provided for 2030 and 2060 as benchmarks. Under the reference scenario, modest declines in $PM_{2.5}$ exposure are accompanied by a sharp increase in mortality, rising to 2.8 million deaths by 2060. In contrast, stringent emission control and climate mitigation pathways lead to substantially cleaner air and fewer deaths, with CleanAir capturing the progressive benefits across scenarios in close agreement with CMAQ. Across all pathways, CleanAir closely reproduces CMAQ estimates of both $PM_{2.5}$ concentrations and health impacts. For national population-weighted mean $PM_{2.5}$ concentration (i.e., $PM_{2.5}$ exposure), the two models agree with R = 0.998, NMB = –2.5%, and RMSE = 0.73 μg $m^{-3}$. For $PM_{2.5}$-related mortality, R = 0.998, NMB = –1.9%, and RMSE = 0.06 million people. Spatial patterns are similarly consistent (Fig. 5c; R = 0.989–0.995), with pollution hotspots concentrated in eastern China, and the greatest improvements occurring in heavily polluted regions under cleaner pathways. These results demonstrate CleanAir's ability to capture both regional disparities and long-term trends in air quality.

What distinguishes CleanAir is its efficiency. Running all long-term scenarios over four decades requires approximately 185 days with CMAQ on a 40-CPU supercomputer, whereas CleanAir can complete the same task in 6 minutes on a single RTX 5090 GPU. This remarkable efficiency enables researchers and policymakers to explore a wind range of emission pathways quickly and cost-effectively, supporting decision-making for long-term air quality and health management.

**Discussions**

In this study, we present CleanAir, a deep learning-based model designed for $PM_{2.5}$ concentration regulation in China by accelerating traditional CTMs, which are often computationally intensive and time-consuming. CleanAir simulates daily, gridded concentrations of $PM_{2.5}$ and its chemical component in emission reduction scenarios with any given meteorological conditions and emission levels lower than its baseline, allowing for 0–100% reduction ratios across five major species ($SO_2$, $NO_x$, primary PM, NMVOCs and $NH_3$) and three emission layers, at a horizontal resolution of 36 km. While CleanAir does not fully replicate all the functions of CMAQ, it demonstrates strong capability in capturing the effects of emission reductions on $PM_{2.5}$ concentrations—at a speed five orders of magnitude faster.

This substantial efficiency gain enables scientists and policymakers to explore a broader range of short- and long-term emission scenarios, facilitating the identification of more effective air quality management strategies.

Despite its strong ability to capture emission-concentration relationships under complex reduction scenarios, the CleanAir model has several limitations. First, it currently simulates only $PM_{2.5}$ and its chemical components, while other key pollutants such as ozone, $NO_2$, $SO_2$, and CO are not included as predicting targets. This limits the model's applicability, particularly in supporting the co-control of multiple pollutants. Second, the current training dataset consists solely of emission reduction scenarios, without sufficient representation of emission increase cases. As a result, CleanAir has limited ability to capture $PM_{2.5}$ concentration responses to rising emissions, reducing its applicability in scenarios involving economic recovery, industrial expansion, or relaxation of emission controls. In addition, applying CleanAir to meteorological conditions beyond the training period still requires a CMAQ-simulated concentration under the 2017 MEIC-HR emissions for the corresponding meteorological year as input. Although we have pre-simulated such data from 2000 onward to reduce the burden on end users, the model remains partially dependent on CMAQ outputs. Future efforts should focus on reducing this dependency to enhance model independence and usability.

Deep learning-based air quality prediction remains in its early stages but holds significant potential for future advancement[38]. Improvements may focus on both enhancing simulation capabilities and improving modeling methodologies. As for simulation capability, expanding CleanAir's scope to incorporate more air pollutants would enable more comprehensive assessment of air quality regulations. Integrating real-world observational data alongside CMAQ simulations could help reduce model uncertainty. Furthermore, extending the training dataset to include emission increase scenarios would broaden the model's applicability in diverse regulatory contexts. On the methodological side, incorporating physics- or chemistry-informed deep learning approaches[39] may improve model interpretability and reliability, helping to open the "black box" of predictions. As AI technologies continue to evolve, such advancements are expected to drive the next generation of deep learning-based air quality prediction, offering more efficient and accurate tools to support environmental decision-making and policy evaluation.

## Acknowledgements

This work was supported by the National Natural Science Foundation of China (42222507 to G.G.) and the New Cornerstone Science Foundation through the XPLORER PRIZE to Q.Z.

## Author Contributions

Q.Z., G.G., and X.H. conceived the study. S.L. performed the research. Y.X. helped with coding the CleanAir model. H.H. helped with data analysis. S.L., G.G., Q.Z., and X.H. interpreted data. S.L., G.G., Q.Z., and X.H. wrote the paper with input from all co-authors.

## Competing Interests

A patent application related to this work has been submitted on April 29, 2025 (Application No. 2025105546852; Applicant: Tsinghua University; Inventors: Q.Z., G.G., X.H., S.L., and H.H.). The patent covers the methodology for developing the CleanAir model. The remaining authors declare no competing interests.

## Methods

### Preparation of a CMAQ-based training dataset

Unlike deep learning-based meteorological models that are typically developed based on open-access reanalysis datasets, CleanAir was trained and evaluated on a CMAQ-based dataset comprising baseline and a wide range of emission reduction scenarios. This section gives an overview of the CMAQ model used in this study, along with the design of baseline and emission reduction scenarios. In addtion, we describe the approach to randomly splitting the dataset into training, validation, and test subsets.

*Description of CMAQ and baseline scenario*

CTMs simulate air pollutant concentrations under complex emissions and meteorological conditions. In this study, we applied WRF version 3.9 (https://www.mmm.ucar.edu/models/wrf) and CMAQ version 5.2 (https://github.com/USEPA/CMAQ/tree/5.2) to simulate the $PM_{2.5}$ concentrations over China, where WRF provided meteorological variables as inputs for CMAQ. As an widely used model in $PM_{2.5}$ concentration regulation[7,40], CMAQ has been demonstrated to well reproduce $PM_{2.5}$ spatiotemporal patterns over China. While CMAQ was selected as a representative implementation in this study, the methodological framework for developing CleanAir can be applied to other Eulerian air quality models as well.

The WRF-CMAQ modeling system was configured following Zhang et al. (2019)[41]. The simulation domain (Supplementary Figure 1) covers the entire China with a horizontal resolution of 36 km. For the WRF model, data from the National Centers for Environmental Prediction Final Analysis (NCEP-FNL) were used as initial and boundary conditions, whereas the National Oceanic and Atmospheric Administration (NOAA) 0.25° daily Optimum Interpolation Sea Surface Temperature (OISST) reanalysis data and NCEP Automated Data Processing global observational weather data were used for analysis, observation, and soil nudging. Parameterization schemes followed the setup in Zhang et al. (2019). For the CMAQ model, we used the CB05 gas-phase mechanism (with version 5.1 updates) and sixth-generation CMAQ aerosol mechanism (AERO6). Chemical boundary conditions were provided by CMAQ's default profile. Sea-salt and dust emission were calculated online in the CMAQ model. Biogenic emission were estimated by Model of Emissions of Gases and Aerosols from Nature (MEGAN) version 2.1[29], driven by WRF-simulated meteorology.

When developing the CleanAir model, we selected the year 2017 as baseline, which enables CleanAir to simulate scenarios with emissions equal to or lower than 2017 levels. Anthropogenic emissions for mainland China in 2017 were retrieved from MEIC-HR (http://meicmodel.org.cn)[28]. We considered eight major pollutants: primary $PM_{2.5}$ (7.6 Tg), black carbon (BC; 1.3 Tg), organic carbon (OC; 2.1 Tg), $PM_{10}$ (10.2 Tg), sulfur dioxide ($SO_2$; 10.5 Tg), nitrogen oxides ($NO_x$; 22.0 Tg), ammonia ($NH_3$; 10.3 Tg), and nonmethane volatile organic compounds (NMVOCs; 28.5 Tg). Sectoral contributions to each pollutant are shown in Supplementary Table 2. Outside mainland China, anthropogenic emissions were derived from the MIX Asian emission inventory and fixed at the 2010 levels. Although this assumption may introduce some bias in our simulated $PM_{2.5}$ concentrations, its influence is considered negligible since local emissions dominate $PM_{2.5}$ exposure in mainland China (e.g., 96.5% reported by Zhang et al. 2017[42]). To minimize the influence of initial conditions on simulation results, a 16-day spin-up period was applied prior to the baseline simulation.

*Evaluation of baseline scenario*

Meteorological variables simulated by the WRF model were evaluated against ground-level observations from the National Climate Data Center (NCDC). As shown in Supplementary Table 4, the WRF reproduced surface temperature and relative humidity with high correlation (R=0.97 and 0.73, respectively) and low bias (NMB=-5.14% and 2.29%, respectively; see *Evaluation Methods* for definitions of statistical metrics). Surface wind speed and precipitation were slightly overestimated, with NMBs of 15.96% and 7.38%, respectively, which may partly explain the underestimation of simulated $PM_{2.5}$ concentrations. According to the performance benchmarks proposed by Emery et al. (2017)[34], the WRF simulation is acceptable for the following air quality modeling.

We used ground-level $PM_{2.5}$ observations from the China National Environmental Monitoring Centre (CNEMC; https://air.cnemc.cn:18007; blue dots in Supplementary Figure 1) to evaluate the performance of the CMAQ-simulated baseline $PM_{2.5}$ concentrations. Supplementary Figure 7 compares simulated annual mean $PM_{2.5}$ concentrations with observations across China with continuous data throughout the year. The model successfully captured the national spatial distribution of $PM_{2.5}$ over China with R of 0.61 and RMSE of 17.16 $\mu g\ m^{-3}$. As shown in Supplementary Figure 8, CMAQ also reproduced the temporal variation in daily $PM_{2.5}$ concentration variations over China and key regions, with R from 0.73 to 0.94 and RMSE from 8.07 μg $m^{-3}$ to 21.25 μg $m^{-3}$. Supplementary Table 5 summarizes station-level statistics for daily $PM_{2.5}$ concentrations over different regions. CMAQ performed well with R ranging from 0.59 to 0.78 and NMB from -15.56% to 10.04% over different regions.

We further evaluated simulated $PM_{2.5}$ chemical composition against observation data from multiple sources provided by our previous study[7]. The spatial distribution of these stations is illustrated in Supplementary Figure 1 with orange triangles. Overall, the CMAQ-simulated $PM_{2.5}$ components showed reasonable agreement with observations (Supplementary Figure 9), with R ranging from 0.28 to 0.81, NMB within -50.23% to 14.35%, and RMSE between 2.10 and 18.19 μg $m^{-3}$. The model well simulated $NO_3^-$ and $BC$, while underestimated $SO_4^{2-}$ (NMB = -50.23%) and $OM$ (NMB = -46.18%). These biases may be attributable to uncertainties in emission inventories[43] and limitations in the chemical mechanisms of the CMAQ model[44,45], such as the lack of heterogeneous formation pathways for $SO_4^{2-}$ and $OM$. Despite these limitations, the CMAQ provides a important basis for training the CleanAir model to capture the concentration responses of $PM_{2.5}$ components under emission changes.

*Multi-dimensional sampling of emission reduction scenarios*

The emission reduction scenarios simulated by CMAQ constitute the core of the dataset for developing the CleanAir model. These scenarios provide diverse examples of emission perturbations, enabling CleanAir to learn the complex relationships between emission changes and $PM_{2.5}$ concentration responses. We selected January, April, July, and October of 2017 to create emission reduction scenarios, each spanning one month and representing different meteorological and emission condictions across the four seasons. This strategy significantly reduced computational burden while preserving representative variability of emissions and meteorological conditions throughout the year.

In total, 2,416 emission reduction scenarios (see Supplementary Figure 3 for a diagram) were generated based on two complementary sampling strategies: (1) quasi-random sampling via Sobol's algorithm to ensure broad and uniform coverage of the emission perturbation sampling space and (2) targeted scenario design to capture physical and chemical characteristics of interest. Each scenario was defined by 15 emission reduction ratios for each simulation grid, derived from the combination of five major pollutant species ($PM$, $SO_2$, $NO_x$, $NH_3$, and NMVOCs) and three emission layers. Here PM includes primary $PM_{2.5}$, BC, OC, and $PM_{coarse}$, which were assigned a common reduction ratio to simplify the modeling, as end-of-pipe control measures typically affect these species simultaneously. The three emission layers are defined based on emission heights—surface, intermediate (~50 m), and elevated (~280 m)—reflecting the fact that emissions at different heights influence surface-level concentrations differently. Each emission reduction ratio ranged from 0% (no reduction) to 100% (full reduction), and a complete scenario was determined once the 15 emission reduction ratios were assigned to the grids across the domain.

For sampling strategy 1, we applied Sobol's algorithm[31] to sample emission reduction ratios in the 15-dimentional emission reduction space. Sobol's algorithm is a widely used method for generating quasi-random samples in high-dimensional sample space[46,47]. Compared with traditional random sampling methods such as Monte Carlo, Sobol's algorithm provides higher sampling efficiency by uniformly covering the entire 15-dimensional space. Using the "scipy" package (version 1.15.2) in Python, we simply needed to specify the sampling dimension and the desired number of samples to obtain the emission reduction samples for each grid. However, uniform sampling in each dimension can lead to an uneven distribution of total emission reductions, as reported in previous studies[19]. To mitigate this issue, additional margin processing was applied to the samples, following the method described by Xing et al. (2011)[19]. To reflect real-world regulation at different spatial scales, we assigned emission reduction ratios at three representative levels: national, provincial, and grid levels. For example, in cases of sampling at the provincial scale, all grids within a given province were assigned identical reduction ratios in a certain dimension, representing coordinated regional emission control strategies. By means of Sobol's algorithm sampling, we obtained 1,728 emission reduction scenarios, with each scenario spanning one month.

For sampling strategy 2, we designed specialized scenario samples to cover emission reduction scenarios of particular interest that are not well captured by Sobol's algorithm. Two sets of specialized samples were developed. First, 400 scenarios focused on the reductions of $NO_x$ and NMVOCs. These two precursors play a key role in atmospheric oxidation and secondary aerosol formation; varying them together enables the model to better learn the nonlinear $PM_{2.5}$ responses driven by changes in oxidizing capacity. Second, 288 scenarios were constructed to isolate emission reductions from individual emission layers, facilitating the model's ability to distinguish the influence of emission height on surface-level concentrations.

For each emission reduction scenario, the newly generated anthropogenic emissions were incorporated into CMAQ to simulate the corresponding $PM_{2.5}$ concentrations. The model configurations were identical to those used for the baseline scenario (e.g., meteorological conditions, biogenic emissions, etc.), except that the anthropogenic emission inputs were replaced with scenario-specific values. From the CMAQ outputs, we extracted the following variables to construct the training dataset: (1) anthropogenic emissions, including $SO_2$, $NO_x$, $NH_3$, $PM_{2.5}$, BC, OC, $PM_{coarse}$, and NMVOCs across the surface, intermediate, and elevated layers; (2) $PM_{2.5}$ components, including $SO_4^{2-}$, $NO_3^-$, $NH_4^+$, OM, secondary fraction of OM (SOM), BC, $PM_{other}$; and (3) other relevant species, including MDA8 $O_3$, MDA1 $O_3$, and Daily mean $O_3$. In total, 2,416 emission reduction scenarios yielded 74,292 daily samples for CleanAir's development, covering a broad range of emissions and meteorological conditions.

*Training, validation, and test splits*

The CTM-based dataset was divided into training, validation, and test datasets to support the development and evaluation of the CleanAir model. Specifically, 60% of each type of emission reduction scenarios was randomly assigned to the training set, 20% to the validation set for trained model selection (see section *Training strategy* in *The CleanAir model details*), and the remaining 20% to the test set for independent performance evaluation. The split was performed at the scenario level to avoid any data leakage across the three datasets. Each set contains scenarios from all four months (January, April, July, and October) and all spatial sampling scales, ensuring sufficient diversity in meteorological conditions and emission change patterns.

**The CleanAir model details**

This section provides a detailed description of CleanAir, including the model input and outputs, model architecture, the loss function with adaptive weights, and model training strategy.

*Model inputs and outputs*

CleanAir outputs $PM_{2.5}$ concentrations of emission reduction scenarios, with five important inputs: (1) anthropogenic emissions fixed at 2017 levels as baseline; (2) biogenic emissions; (3) meteorological fields; (4) baseline simulations of $PM_{2.5}$ components and other relevant species; and (5) anthropogenic emissions from the emission reduction scenario. Inputs (1)–(4) are simulated in advance using CMAQ based on 2017 MEIC-HR emissions with biogenic emissions and meteorological data varying from 2000 to the present. These inputs serve as default settings embedded in CleanAir without the need for additional simulations when using the model. These inputs and outputs are described in detail below. A schematic illustration is provided in Supplementary Figure 2, and a complete list of variables is documented in Supplementary Table 1.

To represent the unperturbed reference state, CleanAir takes anthropogenic emissions fixed at 2017 levels from the MEIC-HR as the baseline input. The input emissions include eight air pollutant species ($PM_{2.5}$, BC, OC, $PM_{coarse}$, $SO_2$, $NO_x$, $NH_3$, and NMVOCs) across three emission layers within mainland China. For regions outside China, the same eight pollutant species from the MIX emission inventory are included, but the emissions are not disaggregated by sector. Spatially, all emissions are on the CMAQ horizontal grid with 127 rows and 172 columns (i.e., 127 × 172). To meet the max-pooling requirements where the number of grid cells must be divisible by 2 at least twice, the grid is padded to 128 × 176 using reflection padding method. Temporally, they are aggregated as daily totals and span the given day and the previous seven days, as pollutants typically have lifetimes in the atmosphere ranging from several hours to a few days, with most physical and chemical influences diminishing within a week[48]. Those variables are finally concatenated into a data tensor with dimensions of 32 (variables) × 8 (days) × 128 (rows) × 176 (columns).

Biogenic emissions at the time of interest are provided as a background source and are simulated using the MEGAN model driven by WRF meteorological fields. Here, only biogenic NMVOCs is considered, given its importance in the chemical formation of OM. Like anthropogenic emissions, biogenic NMVOCs are input into CleanAir as a data tensor with dimensions of 1 (variable) × 8 (days) × 128 (rows) × 176 (columns).

Meteorological fields at the time of interest are essential for capturing atmospheric physical and chemical processes and are derived from WRF simulations. These inputs comprise a diverse set of 2D and 3D variables with complex spatial structures and grid configurations. For 2D variables in grid system of 127 × 172, 21 variables are included, covering pressure, temperature, humidity, radiation, wind, precipitation, cloud, layer height, heat flux, aerodynamic parameters, and properties of underlying surface. For 3D variables in the same grid system, six variables are considered. These variables are defined across seven vertical layers in CMAQ, approximately spanning from the surface to the top of the planetary boundary layer. To unify input dimensions, each 3D variable is separated by layer and treated as seven different input channels. As a result, they are processed into a data tensor with 42 variables. Additionally, the U- and V-components of true wind are taken from a different grid system (128 × 173). These are also 3D variables across the same seven vertical layers, forming 14 input channels. They are processed by CleanAir's head layers and adjusted to match variables from grid of 127 × 172.

To represent the chemical environment under baseline conditions, CleanAir also uses pre-simulated concentrations of $PM_{2.5}$ components and other relevant species from CMAQ as part of its input. The $PM_{2.5}$ components include seven near-surface species, $SO_4^{2-}$, $NO_3^-$, $NH_4^+$, $OM$, $SOM$, and $BC$, $PM_{other}$. Among other relevant species, three surface ozone indicators—MDA8 $O_3$, MDA1 $O_3$, and Daily mean $O_3$—are defined on the 127 × 172 grid system. In addition, eight 3D reactive intermediates are extracted from the same grid system, each resolved across seven vertical layers, resulting in 56 input variables.

Anthropogenic emissions under emission reduction scenarios, derived from the MEIC-HR or MEIC, are provided as a distinct input to represent perturbations relative to the baseline. Here anthropogenic emissions include $PM_{2.5}$, BC, OC, $PM_{coarse}$, $SO_2$, $NO_x$, $NH_3$, and NMVOCs across three emission layers within mainland China. The input is structured the same as those in the baseline scenario.

The output of CleanAir consists of concentrations of $PM_{2.5}$ components and other relevant species under the emission reduction scenario. Unlike the input variables, which span eight consecutive days, the output variables correspond only to the target day of simulation. The $PM_{2.5}$ components involve the same seven species as inputs, while other relevant species consider three surface ozone indicators (MDA8 $O_3$, MDA1 $O_3$, and Daily mean $O_3$). These outputs are combined into a data tensor with dimensions of 10 (variables) × 1 (day) × 128 (rows) × 176 (columns).

Before being fed into CleanAir, all input and output variables are normalized to a 0 to 1 range using min-max normalization. The minimum and maximum values for each variable are determined based on their distributions in the baseline scenario. This normalization step ensures numerical stability during training and allows the model to process variables with different units and magnitudes in a balanced and consistent manner.

*Model architecture*

CleanAir is built on a Residual Symmetric 3D U-Net[27] architecture (Fig. 1), a variant of the widely used U-Net, chosen for its superior multi-scale feature extraction and fusion capabilities. Given that $PM_{2.5}$ air quality prediction involves complex processes spanning multiple spatiotemporal scales, this architecture is well-suited for capturing these intricate patterns. A detailed description of the model architecture is provided below.

The CleanAir architecture start with head layers, which are designed to align data shapes from different CMAQ grid configurations (i.e., 127 × 172 vs. 128 × 173) and to perform preliminary feature extraction. The head layers consist of two parallel branches: (1) The first branch includes a Conv3D layer followed by a Padding layer. Data from the 128 × 173 grid are fed into this branch, where a convolutional kernel with a shape of (1, 2, 2) and stride of (1, 1, 1) is used to reduce the grid size to 127 × 172. The following Padding layer then adjusts the grid to 128 × 176 to meet the max-pooling requirement; (2) The second branch consists of a single Conv3D layer, where all input data are merged and passed through this layer. The head layers work together to extract features for more advanced feature extraction in subsequent layers of the CleanAir model.

The extracted features are subsequently fed into a symmetric U-Net structure comprising five residual modules with max-pooling and up-sampling layers, as well as skip connections using summation. Each residual module consists of three convolutional blocks made up of Conv3D and GroupNorm layers, each followed by an ELU activation function. GroupNorm is utilized here to normalize features within groups of channels (four groups in this study), which helps stabilize and accelerate training. The ELU activation function allows for small negative values, which can improve learning dynamics, particularly in deeper networks by mitigating issues related to vanishing gradients. A skip connection is implemented to connect the outputs of the first convolutional block with those of the final convolutional block. Between the first, second, and third residual modules, two max-pooling layers with a pool kernel size of (2, 2, 2) are applied, which reduces the spatial size by half, enabling the network to capture increasingly abstract features. Between the third, fourth, and fifth residual modules, two up-sampling layers based on 3D transposed convolution operators are implemented, with kernel size of (3, 3, 3), stride of (2, 2, 2), and padding of (1, 1, 1), effectively recovering the spatial resolution and generating high-level features for output. Skip connections using summation joins are used to connect low-level features and high-level features, facilitating more efficient feature information flow.

After steps described above, the features are fed into tail layers to generate the final outputs of pollutant concentration changes relative to the baseline scenario. The input features, with dimensions of 32 (channels) × 8 (days) × 128 (rows) × 176 (columns), are first passed through a Conv3D layer with a kernel size of (1, 1, 1), reducing the channels to 10. Subsequently, the first two dimensions (channels and days) are fused into a single dimension, reshaping the 3D features into 2D features. These 2D features are then processed by multiple Conv2D layers with ReLU activation, and the output tensor has dimensions of 10 (channels) × 128 (rows) × 176 (columns), where the 10 channels represent the concentration changes of seven $PM_{2.5}$ components and three surface ozone indicators.

The final concentration changes for the seven $PM_{2.5}$ components and three surface ozone indicators are then added to their respective baseline values to generate the emission reduction scenario values.

*The adaptive weighted loss function*

In the loss function for CleanAir training, we accounted for both absolute concentrations and concentration changes relative to the baseline scenario for $PM_{2.5}$ components and other reactants. However, this posed a significant challenge in designing the loss function, as multiple components required appropriate weighting, which is important to model performance and trainability. Since manually searching for optimal weights is prohibitive, we incorporated an adaptive weighted loss function inspired by multi-task learning studies[32,33] to address this issue. A conceptual diagram of the adaptive weighted loss function is shown in Supplementary Figure 4.

The total loss function ($\mathrm{L_{total}}$) consists of three terms, as shown below: (1) a loss function that evaluates the accuracy of concentration estimation task ($\mathrm{L_{conc}}$); (2) a loss function that evaluates the accuracy of concentration change estimation task ($\mathrm{L_{\Delta conc}}$); and 3) a logarithmic regularization term that prevents the weighting factors from becoming excessively small.

$$L_{total} = \frac{1}{2(\sigma_{conc}^2 + \varepsilon^2)} L_{conc} + \frac{1}{2(\sigma_{\Delta conc}^2 + \varepsilon^2)} L_{\Delta conc} + \ln(1 + \sigma_{conc}^2)(1 + \sigma_{\Delta conc}^2)$$

The subscripts "conc" and "Δconc" denote the tasks of concentration estimation and concentration change estimation, respectively. $\sigma$ is an adaptive weight automatically determined and adjusted by the training algorithm, while $\varepsilon$ is a small constant (set to 0.01 in this study) used as a constraint to prevent the denominator from approaching zero.

The specific definitions of $L_{conc}$ and $L_{\Delta conc}$ are described below, where the subscript i ranges from 1 to 10, representing the seven $PM_{2.5}$ components and three other relevant species: $SO_4^{2-}$, $NO_3^-$, $NH_4^+$, OM, SOM, BC, $PM_{other}$, MDA8 $O_3$, MDA1 $O_3$, Daily mean $O_3$.

$$L_{conc} = \sum_{i=1}^{10} \left[ \frac{1}{2(\sigma_i^2 + \varepsilon^2)} L_{i,conc} + \ln(1 + \sigma_i^2) \right]$$

$$L_{\Delta conc} = \sum_{i=1}^{10} \left[ \frac{1}{2(\sigma_i^2 + \varepsilon^2)} L_{i,\Delta\, conc} + \ln(1 + \sigma_i^2) \right]$$

The mathematical forms and intrinsic meanings of $L_{conc}$ and $L_{\Delta conc}$ are similar to those of $L_{total}$; here, they treat the estimations of different $PM_{2.5}$ components as separate tasks, with each task having its own corresponding loss function ($L_{i,conc}$ and $L_{i,\Delta\, conc}$). We define $L_{i,conc}$ as follows:

$$L_{i,conc} = \frac{\sqrt{\sum_{m=1}^{M} \sum_{n=1}^{N} \left(C_{m,n}^{CleanAir} - C_{m,n}^{CMAQ}\right)^2}}{\sqrt{\sum_{m=1}^{M} \sum_{n=1}^{N} \left(C_{m,n}^{CMAQ}\right)^2}}$$

where C represents concentrations, subscripts m and n denote row and column indices, respectively, within the gridded simulation domain, with M and N representing the total number of rows and columns. The superscripts CleanAir and CMAQ indicate the data source.

As for $L_{i,\Delta\, conc}$, it is defined as follows:

$$L_{i,\Delta\, conc} = \sqrt{\sum_{m=1}^{M} \sum_{n=1}^{N} V_{m,n} \left(\Delta C_{m,n}^{CleanAir} - \Delta C_{m,n}^{CMAQ}\right)^2}$$

among which the weights $V_{m,n}$ are calculated separately for each grid:

$$V_{m,n} = \frac{\left|\Delta C_{m,n}^{CMAQ}\right|}{\sum_{m=1}^{M} \sum_{n=1}^{N} \left|\Delta C_{m,n}^{CMAQ}\right|}$$

They were determined based on concentration change values, assigning greater weights to grid cells with larger concentration changes. This property allows the training algorithm to prioritize areas with significant concentration changes, thereby improving estimation performance.

Overall, we treated the estimations as separate tasks and designed a loss function that incorporates multi-task learning. The key components of the loss function are the adaptive weights, which enables the training algorithm to automatically balance the contributions of different $PM_{2.5}$ components and other reactants, making the training process more adaptive and efficient.

*Training strategy*

The CleanAir model was trained following a distributed training strategy inspired by FourCastNet[21], with reference of more details to its publicly available implementation (https://github.com/NVlabs/FourCastNet).

A global batch size of 8 was adopted, with each DCU processing one sample per iteration. Optimization was performed using the Fused Adam optimizer, which supports mixed-precision training. This technique reduces memory usage and accelerates training while maintaining model accuracy. The initial learning rate was set to $5 \times 10^{-4}$, consistent with FourCastNet's configuration. We also employed a cosine annealing learning rate schedule, in which the learning rate gradually decays to zero over the training process, helping to improve convergence and generalization.

The model was trained for 16 epochs in total, requiring about 40 hours. To monitor performance and prevent overfitting, the CTM-based dataset was randomly split into training (60%), validation (20%), and test (20%) sets at the scenario level. During each epoch, model parameters were updated using the training set, after which model performance was evaluated on the validation set. Model checkpoints recording the parameters with the best validation performance were iteratively saved and updated, and the checkpoint achieving the highest validation accuracy was ultimately selected as the final trained model.

**Evaluation methods**

We conducted comprehensive evaluations for the CleanAir model against both the CMAQ model and ground-level observations. CMAQ was configured consistently with the methods described in the section *Multi-dimensional sampling of emission reduction scenarios*. This section outlines the evaluation protocol, including definitions of statistical matrics and the performance evaluations in three applications: simulations with different meteorological conditions and emission inventories from 2017 to 2020, short-term emission perturbation scenarios, and long-term emission mitigation pathways.

*Statistical matrices*

The $PM_{2.5}$ concentration from CleanAir and CMAQ from 2017 to 2020 were evaluated against in-situ ground measurements of CNEMC. Statistical metrics considered in our evaluation include Pearson's correlation coefficient (R), mean bias (MB), root-mean-squared error (RMSE), normalized mean bias (NMB), and normalized mean error (NME). Equation for calculating them are documented as follows, with *Sim* represents the CleanAir or CMAQ simulated results and *Obs* means the observed $PM_{2.5}$ concentrations, and *i* represents data samples.

$$R = \frac{\sum_{i=1}^{n}\left((Sim_i - \overline{Sim}) \times (Obs_i - \overline{Obs})\right)}{\sqrt{\sum_{i=1}^{n}(Sim_i - \overline{Sim})^2} \times \sqrt{\sum_{i=1}^{n}(Obs_i - \overline{Obs})^2}}$$

$$MB = \frac{\sum_{i=1}^{n}(Sim_i - Obs_i)}{n}$$

$$RMSE = \sqrt{\frac{\sum_{i=1}^{n}(Sim_i - Obs_i)^2}{n}}$$

$$NMB = \frac{\sum_{i=1}^{n}(Sim_i - Obs_i)}{\sum_{i=1}^{n}Obs_i} \times 100\%$$

$$NME = \frac{\sum_{i=1}^{n}|Sim_i - Obs_i|}{\sum_{i=1}^{n}Obs_i} \times 100\%$$

*Evaluation for MEIC-based simulations from 2017 to 2020*

According to the design of baseline and emission reduction scenarios, CleanAir is applicable to simulating $PM_{2.5}$ concentrations under anthropogenic emissions that are equal to or lower than the 2017 levels. To demonstrate this, we used annually resolved anthropogenic emissions from MEIC (http://meicmodel.org.cn), together with corresponding meteorological conditions, to simulate $PM_{2.5}$ concentrations in China from 2017 to 2020. During this period, China continuously implemented clean air actions, resulting in a steady decline in anthropogenic emissions[7].

The CMAQ model was used to provide reference $PM_{2.5}$ concentrations for performance evaluation. We evaluated CleanAir's performance by comparing it with CMAQ across multiple dimensions: spatial distribution of annual $PM_{2.5}$ concentrations, national population-weighted annual mean $PM_{2.5}$ trends, and daily $PM_{2.5}$ concentrations against ground-level observations. Ground-level observations were obtained from the China National Environmental Monitoring Centre (CNEMC; https://air.cnemc.cn:18007), and population distribution data for 2017–2020 were taken from the Gridded Population of the World, Version 4.11 (GPW v4.11; https://beta.sedac.ciesin.columbia.edu).

*Evaluation for short-term emission perturbation application*

Simulating $PM_{2.5}$ air quality under short-term emission perturbations is widely used in practical $PM_{2.5}$ regulation to support rapid policy assessment. In this study, we designed three emission reduction scenarios of varying control intensities (weak, moderate, and aggressive), targeting a regulation region that includes 57 cities across Beijing, Tianjin, Hebei, Shandong, Shanxi, and Henan (Supplementary Figure 1; Supplementary Table 3). These regions represent China's key area for air pollution prevention and control.

We assigned city-specific emission reduction ratios for each scenario, which were derived from actual emission control practices implemented during the 2022 Beijing Winter Olympics and reflected the spatial heterogeneity in control efforts across cities. The emission reductions focused on four key anthropogenic pollutant species, including $SO_2$, $NO_x$, primary $PM_{2.5}$, and NMVOCs, which are predominantly emitted from power, industry, residential, and transportation sectors. These sectors are typically targeted in short-term emergency controls. $NH_3$, mostly from agriculture sources, remains unchanged due to challenges in short-term regulation. Overall, $SO_2$, $NO_x$, primary $PM_{2.5}$, and NMVOCs within the regulation region were reduced by 11%, 15%, 16%, and 7%, respectively, in weak control scenario; 19%, 33%, 32%, and 24%, respectively, in moderate control scenario; and 43%, 46%, 46%, and 49%, respectively, in aggressive control scenario. All scenarios were simulated for February 2017 with the same emission inventories and meteorological conditions as baseline scenario. This month was selected as it featured two typical regional haze episodes, offering a representative case for evaluating CleanAir's applicability under realistic pollution conditions. Then CMAQ was used to generate the benchmark $PM_{2.5}$ concentrations for comparison.

In this application, we focused on evaluating the model's performance in capturing daily

$PM_{2.5}$ concentration levels across all cities within regulation region, with particular attention to the magnitude of peak $PM_{2.5}$ concentrations. These pollution indicators are critical for assessing the effectiveness of short-term emission controls, as they directly influence air quality classifications and the severity of public health risks during pollution episodes.

*Evaluation for long-term emission mitigation application*

Simulating $PM_{2.5}$ concentrations under future emission mitigation pathways is important for supporting long-term air quality management and reducing the health burden of air pollution. In this study, we used a suite of emission mitigation scenarios from the DPEC v1.2 (http://meicmodel.org.cn)[37]. This scenario set integrates varying degrees of air pollution control and carbon mitigation strategies, yielding five distinct pathways: reference, clean air, on-time peak-clean air, on-time peak-net zero-clean air, and early peak-net zero-clean air. These pathways reflect progressively stronger policy interventions and highlight potential synergies between air quality improvements and climate goals.

In this application, we focused on evaluating the $PM_{2.5}$ exporsure and the associated long-term health impact. $PM_{2.5}$ exporsure is directly linked to population-level health risks and is a key indicator in environmental health policy. Long-term health impacts, such as $PM_{2.5}$-related premature mortality, account for both exposure intensity and duration. To reduce systematic model biases, CleanAir and CMAQ simulations were calibrated against observation-based hindcast $PM_{2.5}$ concentrations from the Tracking Air Pollution in China (TAP) dataset as in the previous study[37].

We used exposure-response function in the Global Burden of Disease (GBD) 2019 study (http://ghdx.healthdata.org/gbd-2019) to quantify the $PM_{2.5}$-related mortality. Six causes of mortality were considered, including chronic obstructive pulmonary disease (COPD), ischemic heart disease (IHD), acute lower respiratory infection (LRI), type II diabetes (DM), lung cancer (LC), and stroke.

The relative risk for $PM_{2.5}$ exposure was modeled using meta-regression-Bayesian, regularized, trimmed (MR-BRT) splines as follows:

$$\mathrm{RR(C)} = \begin{cases} 1, & \mathrm{C} \leq \mathrm{tmrel} \\ \dfrac{\mathrm{MRBRT(C)}}{\mathrm{MRBRT(tmrel)}}, & \mathrm{C} > \mathrm{tmrel} \end{cases} \quad (407)$$

where RR represents the relative risk, C denotes $PM_{2.5}$ concentration exposure, and tmrel is the theoretical minimum-risk exposure level. Based on the relative risk, the $PM_{2.5}$-related mortalities were calculated as follows:

$$M_i = P_i \times PS_a \times B_a \times \frac{RR_{i,a}(C_i) - 1}{RR_{i,a}(C_i)}$$

where i and a denote the grid index and the age group, respectively; M is the estimated long-term $PM_{2.5}$-related mortality; P is the total population; PS is the age structure of the population; and B represents the baseline mortality rate. Historical baseline mortality rates were obtained from the GBD 2019 database, while future estimates were projected according to the World Population Prospects 2019 report[49].

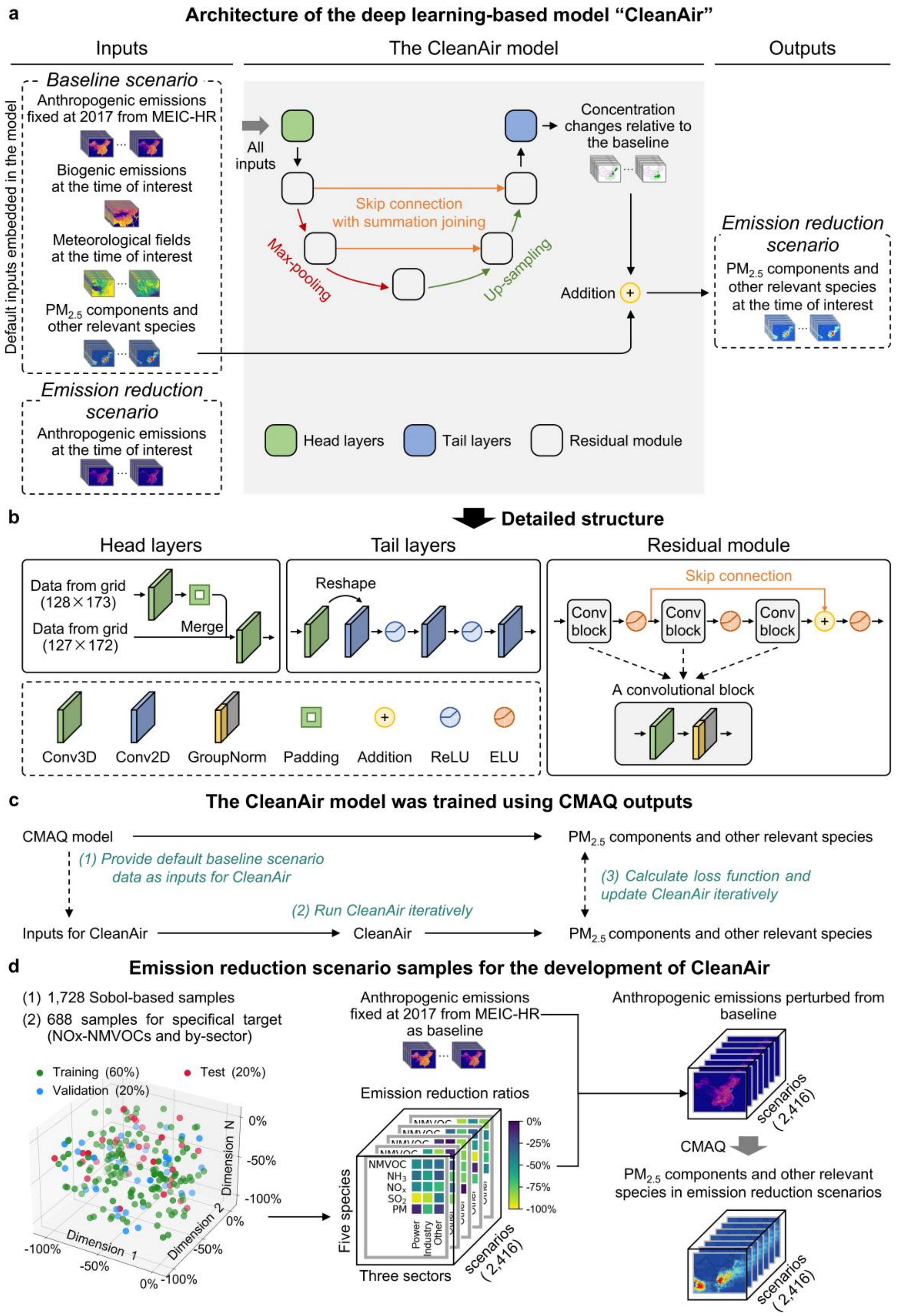

**Fig. 1 | Architecture and training process of the CleanAir model. a,** Overall architecture of the CleanAir model. **b,** Detailed structure of head layers, tail layers, and residual module. **c,** Training process diagram using CMAQ outputs. **d,** Preparation of emission reduction scenario samples for the development of CleanAir.

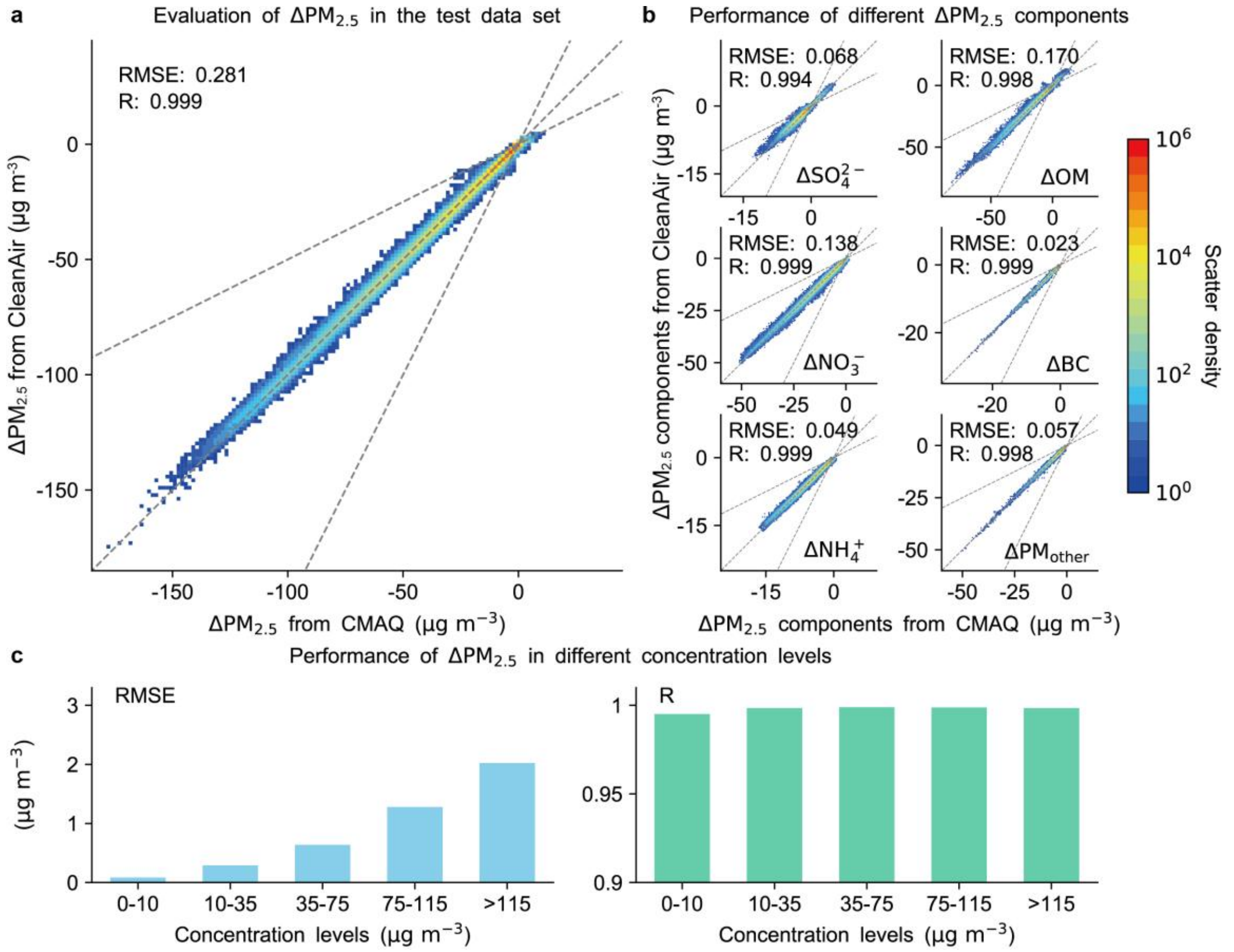

**Fig. 2 | Evaluation of the CleanAir model on the test dataset. a,** Evaluation of monthly averaged $\Delta PM_{2.5}$ (unit: $\mu g\ m^{-3}$), where the x-axis and y-axis represent CMAQ and CleanAir, respectively. **b,** Performance of monthly averaged different $\Delta PM_{2.5}$ components. **c,** Performance of monthly averaged $\Delta PM_{2.5}$ in different concentration levels (0–10 $\mu g\ m^{-3}$, 10–35 $\mu g\ m^{-3}$, 35–75 $\mu g\ m^{-3}$, 75–115 $\mu g\ m^{-3}$, and >115 $\mu g\ m^{-3}$). It should be noted that $\Delta$ represents concentration changes relative to the baseline scenario, R represents Pearson's correlation coefficient, and RMSE represents root mean square error. The gray dashed lines in **a** and **b** are 1:1, 1:2, and 2:1 lines. Evaluation results at the daily average scale are shown in Supplementary Figure 5.

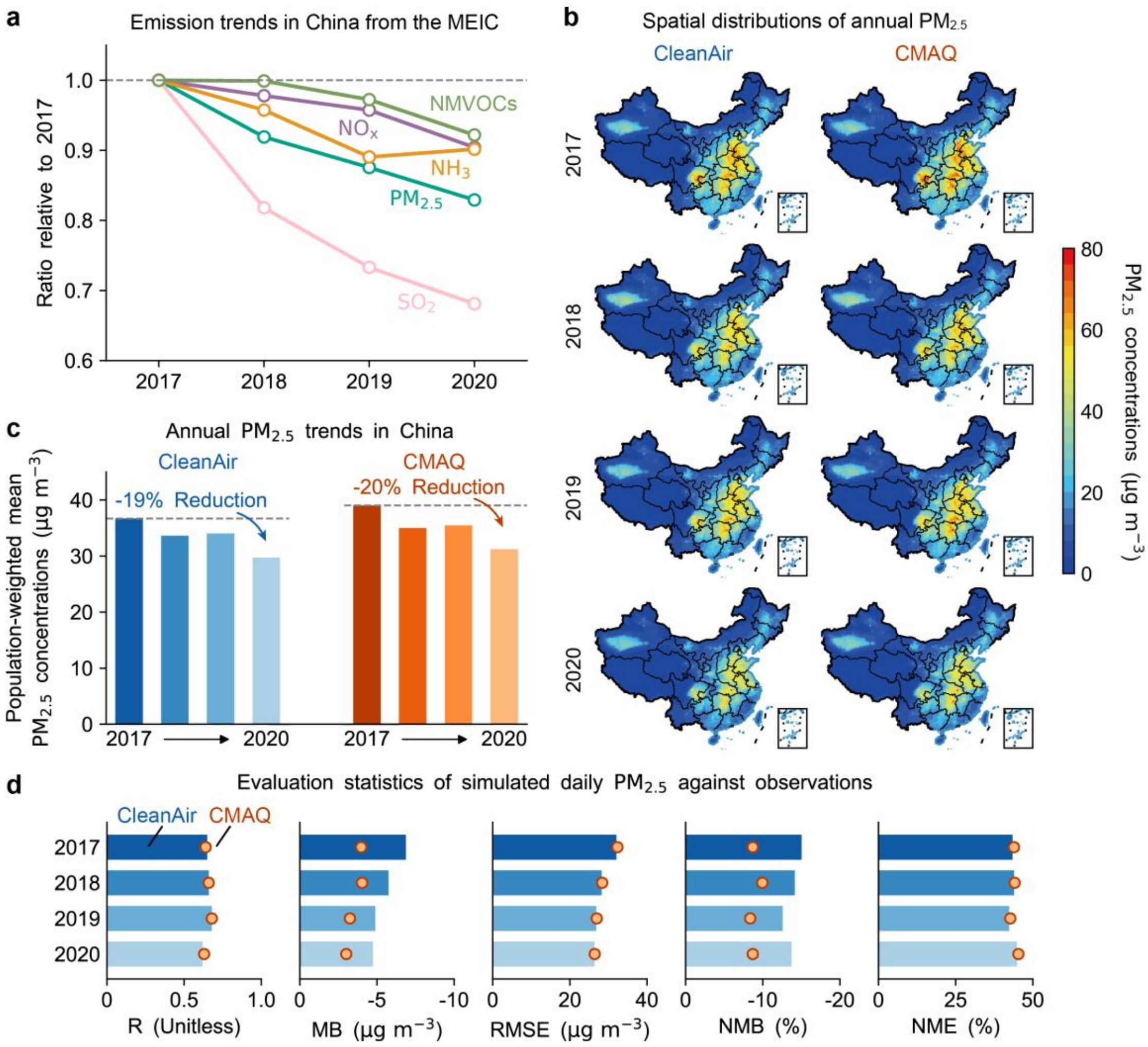

**Fig. 3 | Evaluation for MEIC-based simulations from 2017 to 2020 between CleanAir, CMAQ, and ground-level observations. a,** Anthropogenic emission trends in China from 2017 to 2020 for five major pollutants ($PM_{2.5}$, $SO_2$, $NO_x$, $NH_3$, and NMVOCs) from the MEIC. **b,** Spatial distributions of annual $PM_{2.5}$ concentrations. **c,** Annual population-weighted mean $PM_{2.5}$ trends in China. **d,** Evaluation statistics of simulated daily $PM_{2.5}$ from CleanAir and CMAQ against observations, with statistical metrics defined in *Methods*.

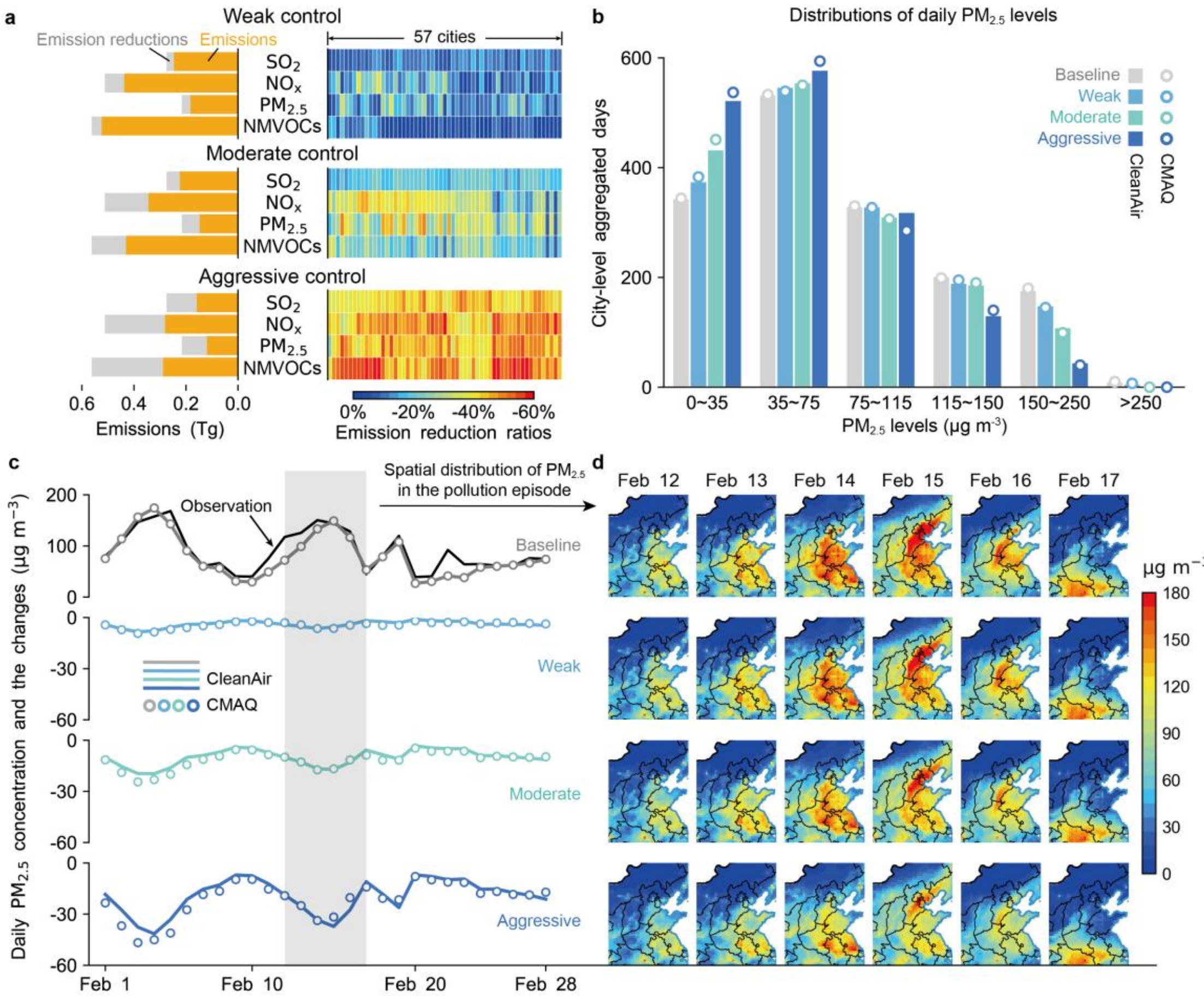

**Fig. 4 | Evaluation for short-term emission perturbation application in February 2017 between CleanAir and CMAQ. a,** Emissions and emission reductions of three emission reduction scenarios relative to the baseline scenario, with heterogeneous emission reduction ratios across 57 cities in the regulatory region (Supplementary Figure 1 and Supplementary Table 3). **b,** Distribution of daily $PM_{2.5}$ levels under the baseline and three emission reduction scenarios, aggregated across 57 cities. **c,** Daily $PM_{2.5}$ concentration and the changes annotated with baseline peak values, along with **d,** an example spatial distribution of $PM_{2.5}$ in the pollution episode from February 12 to 17 simulated by CleanAir.

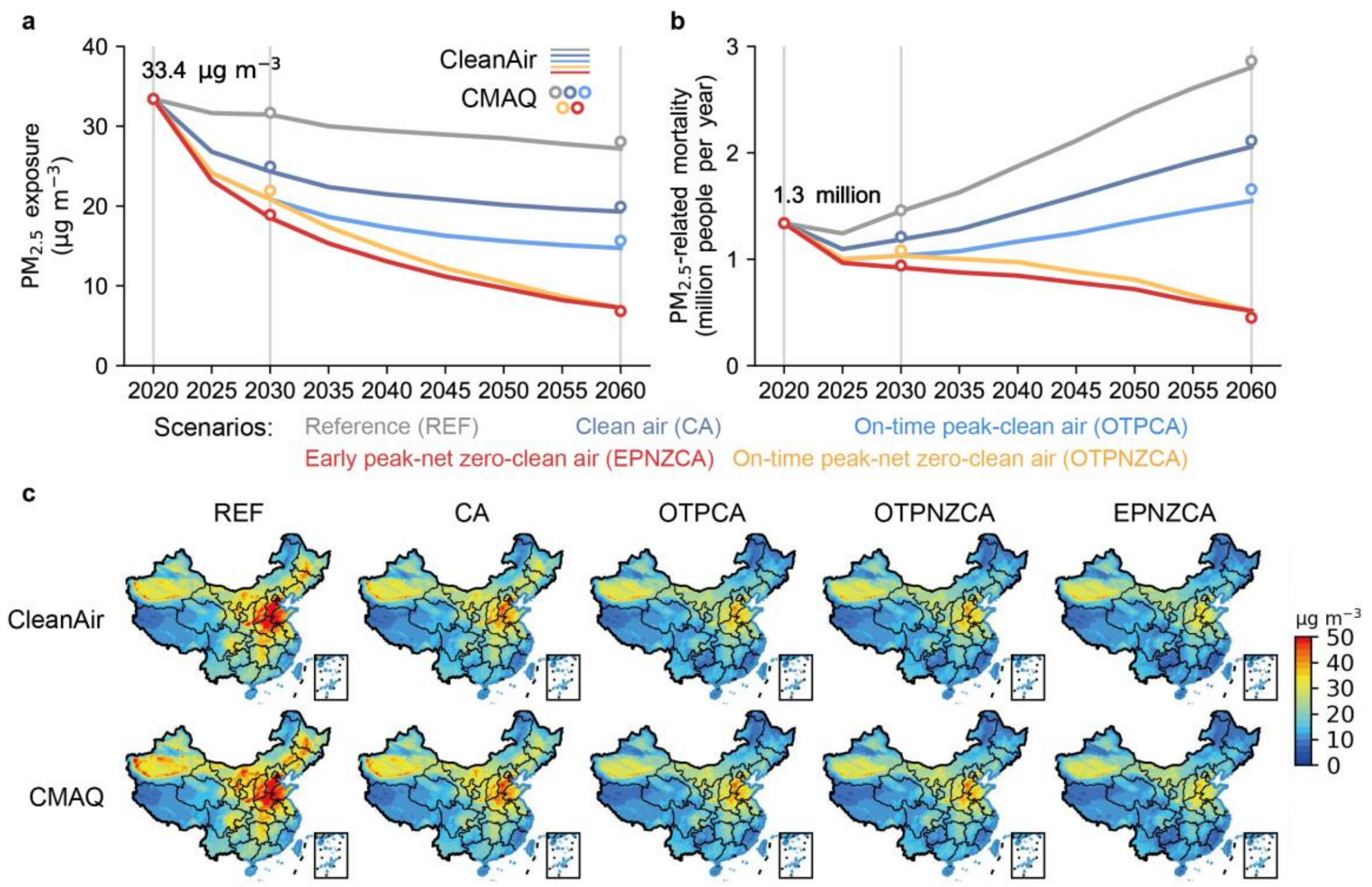

**Fig. 5 | Evaluation for long-term emission mitigation application from 2020 to 2060 between CleanAir and CMAQ. a,** Trajectories of China's $PM_{2.5}$ exposure under different scenarios. **b,** Trajectories of China's $PM_{2.5}$-related mortality under different scenarios. **c,** Spatial distribution of annual $PM_{2.5}$ concentrations in 2030.

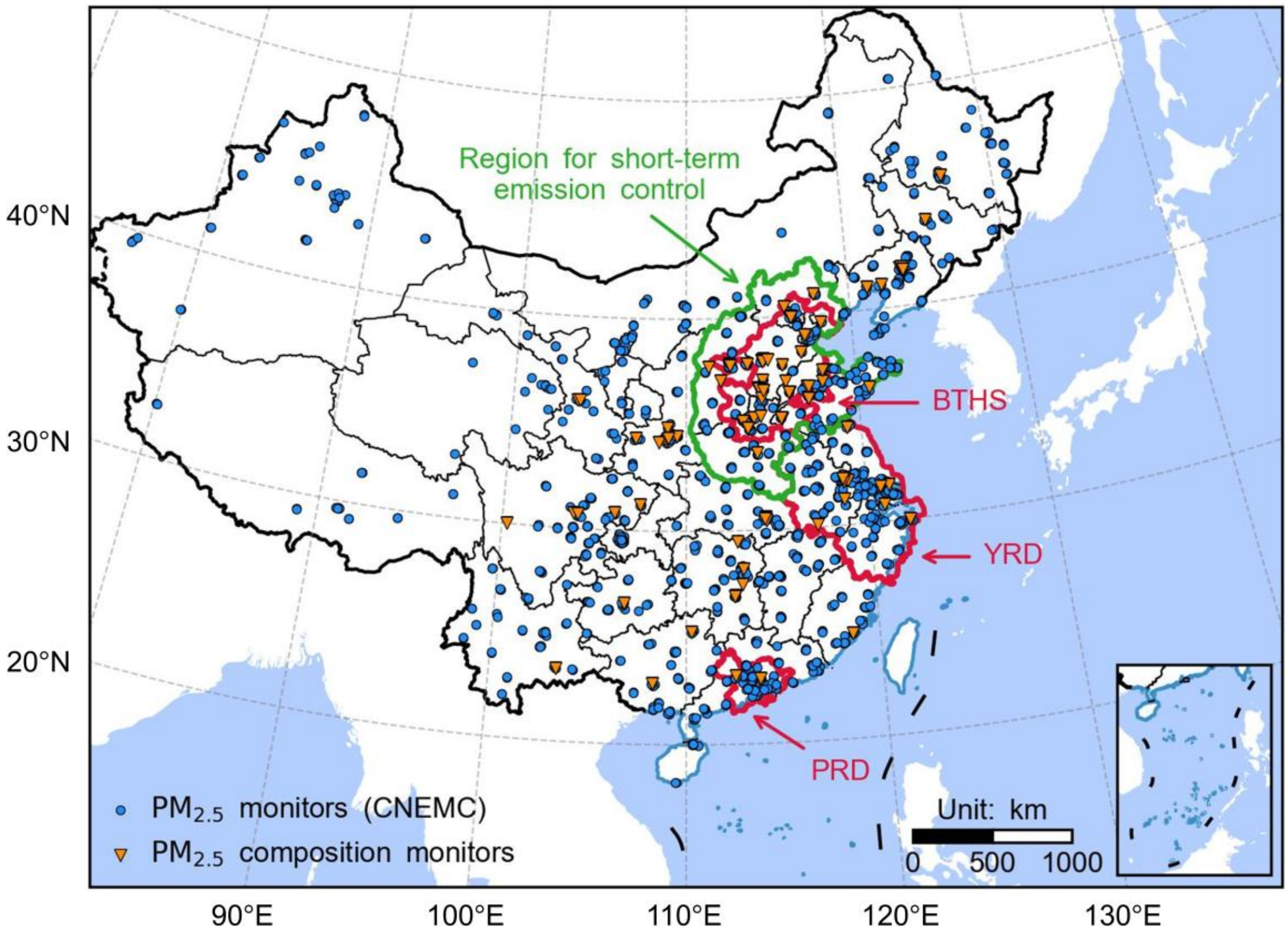

**Supplementary Figure 1 | Simulation domain of CleanAir and CMAQ, overlaid with monitoring sites used in this study and definitions of key regions.** Blue circles represent the locations of CNEMC sites with continuous $PM_{2.5}$ measurements from 2017 to 2020. Orange triangles indicate locations of collected $PM_{2.5}$ chemical composition measures. Red lines outline three key regions, including Beijing-Tianjin-Hebei and its surrounding areas (BTHS), the Yangtze River Delta (YRD), and the Pearl River Delta (PRD). The green line outlines the region for short-term emission control, covering 57 cities in total.

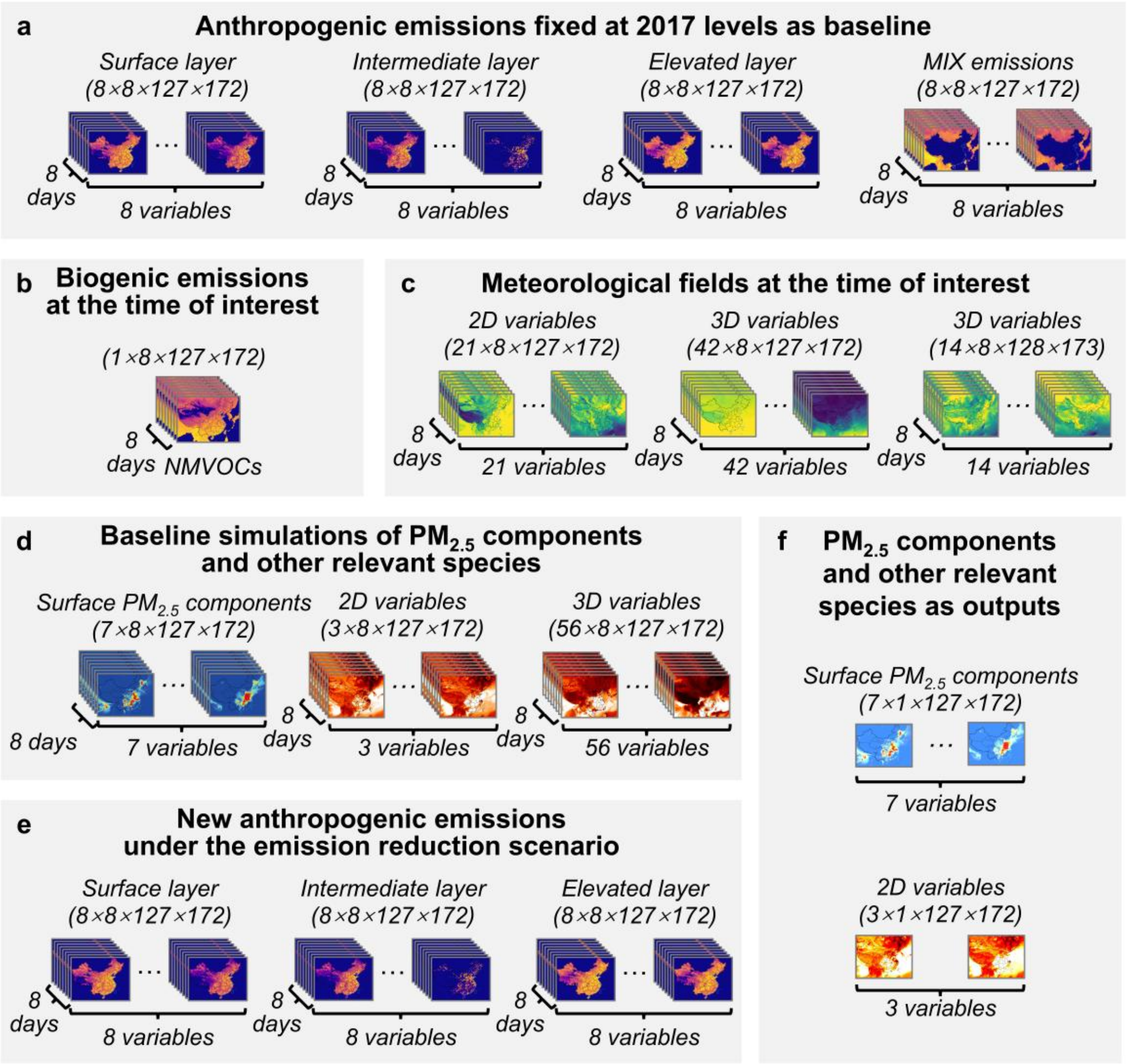

**Supplementary Figure 2 | Diagram of the model inputs and outputs of CleanAir.** Summary of variables and their shapes before padding, including (**a**) anthropogenic emissions fixed at 2017 levels used as the baseline, (**b**) biogenic emissions at the time of interest, (**c**) meteorological fields at the time of interest, (**d**) baseline simulations of $PM_{2.5}$ components and other relevant species, (e) new anthropogenic emissions under the emission reduction scenario, and (**f**) $PM_{2.5}$ components and other relevant species as outputs. Inputs are (a)–(e), and the output is (f).

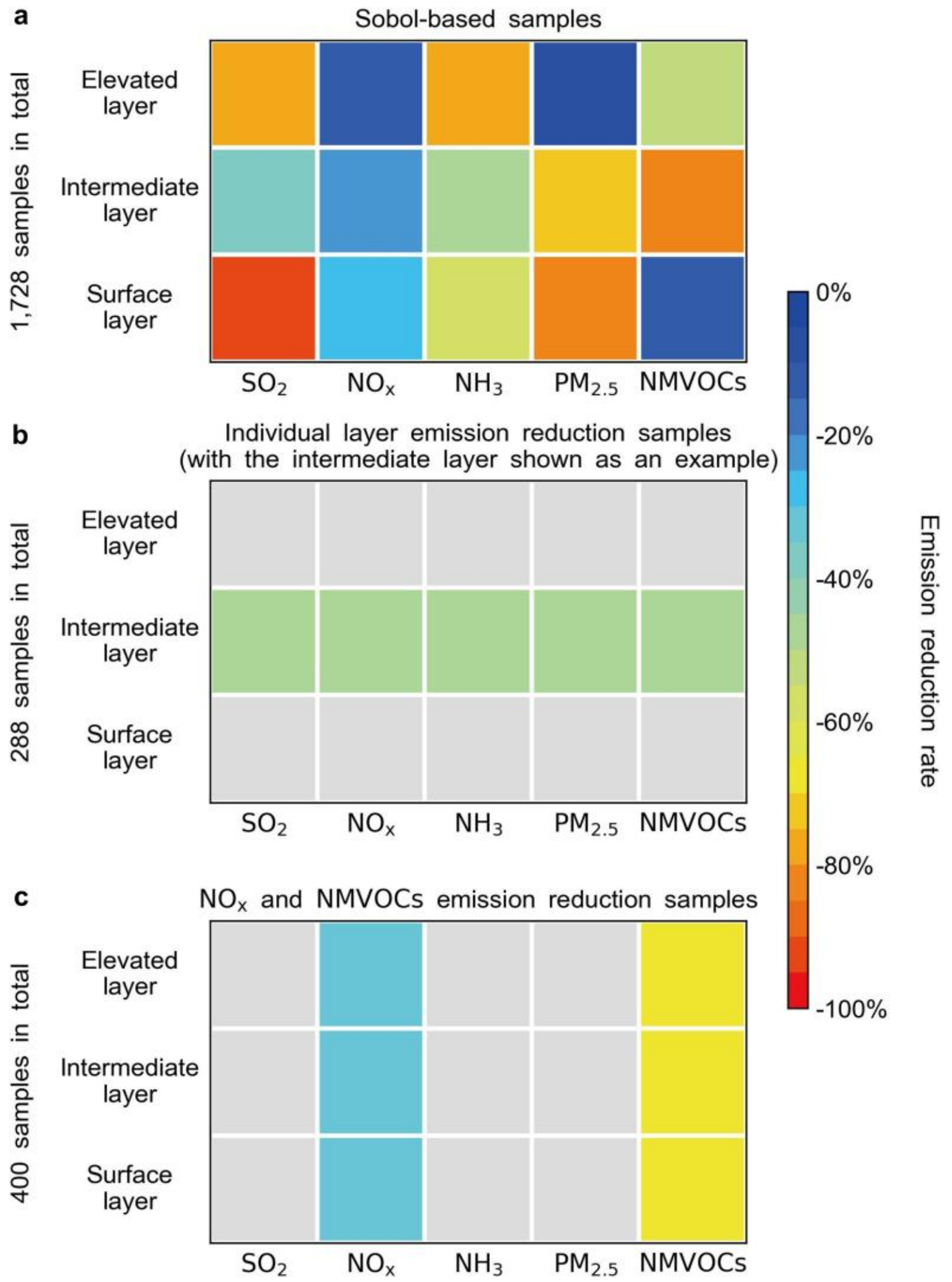

**Supplementary Figure 3 | Diagram of the training dataset containing 2,416 emission reduction scenarios.** Emission reduction ratio examples of (**a**) Sobol-based samples, (**b**) emission reduction samples of individual layer (with the intermediate layer shown as an example), and (**c**) $NO_x$ and NMVOCs emission reduction samples.

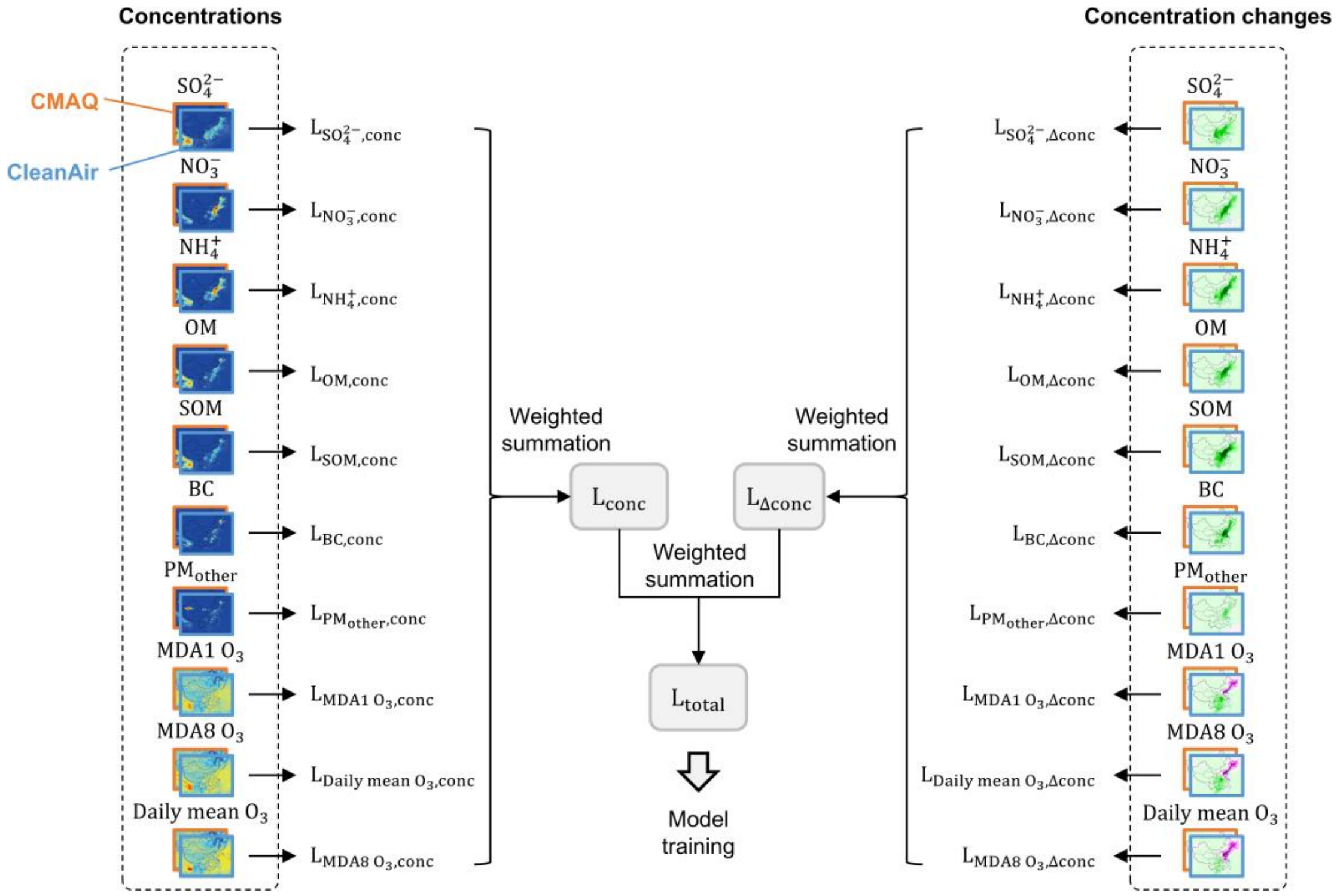

**Supplementary Figure 4 | A conceptual diagram of the adaptive weighted loss function incorporating multi-task learning considerations.** The mathematical definition of loss functions is described in *Methods*.

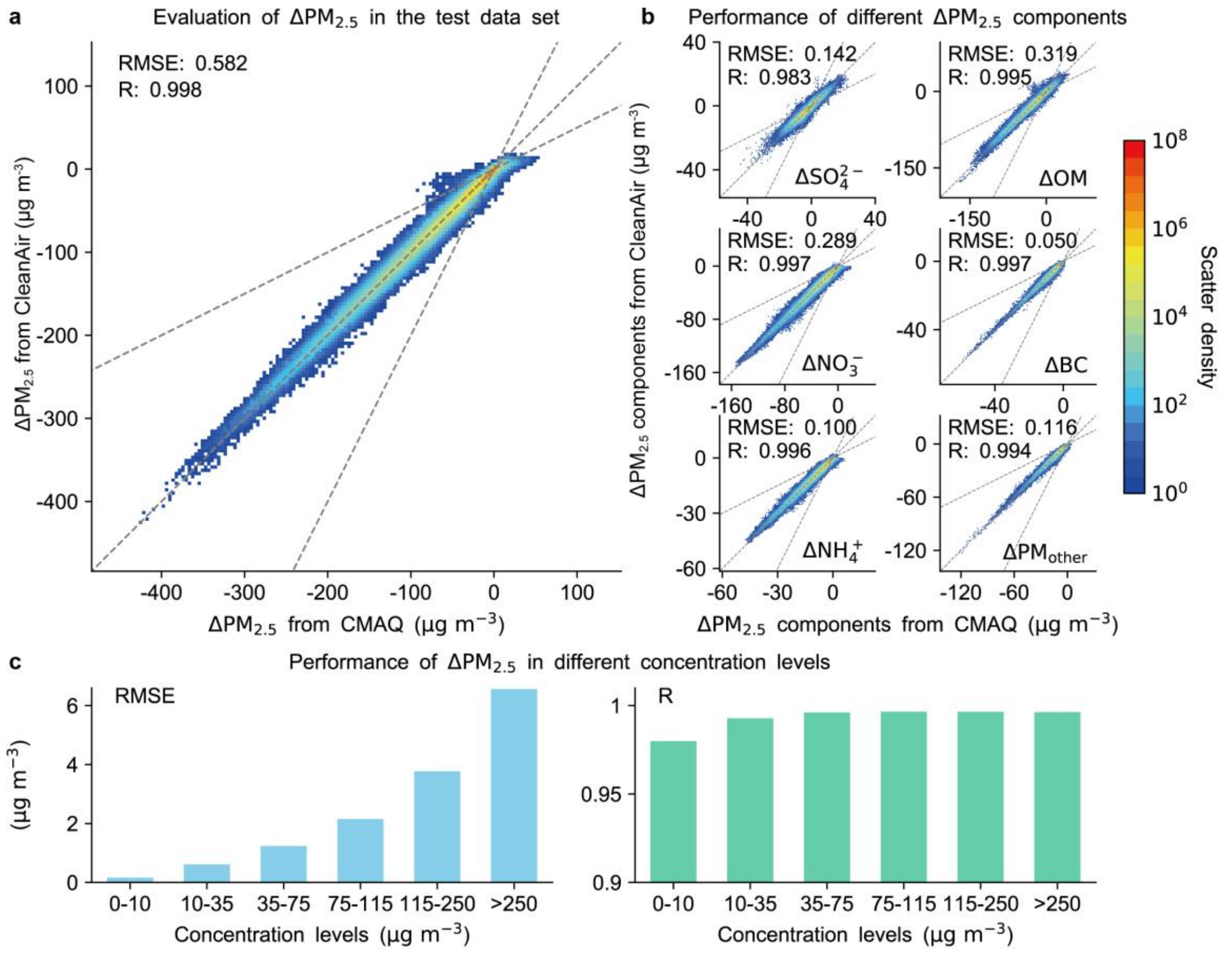

**Supplementary Figure 5 | The same as Fig. 2 but at the daily average scale.**

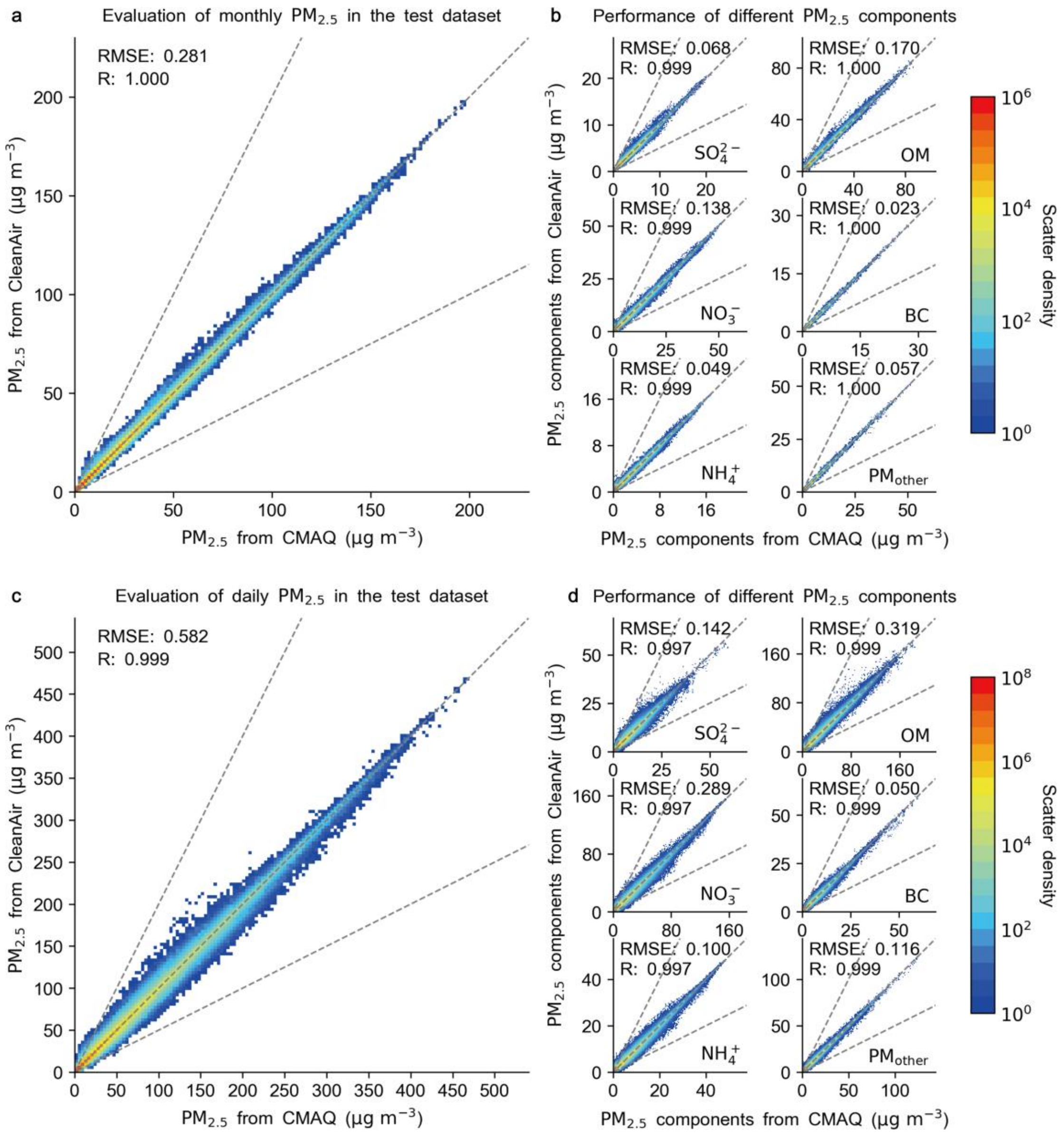

**Supplementary Figure 6 | Evaluation of CleanAir-simulated $PM_{2.5}$ and its component concentrations against CMAQ in the test dataset.** Performance of monthly $PM_{2.5}$ (**a**) and its components (**b**). Performance of daily $PM_{2.5}$ (**c**) and its components (**d**).

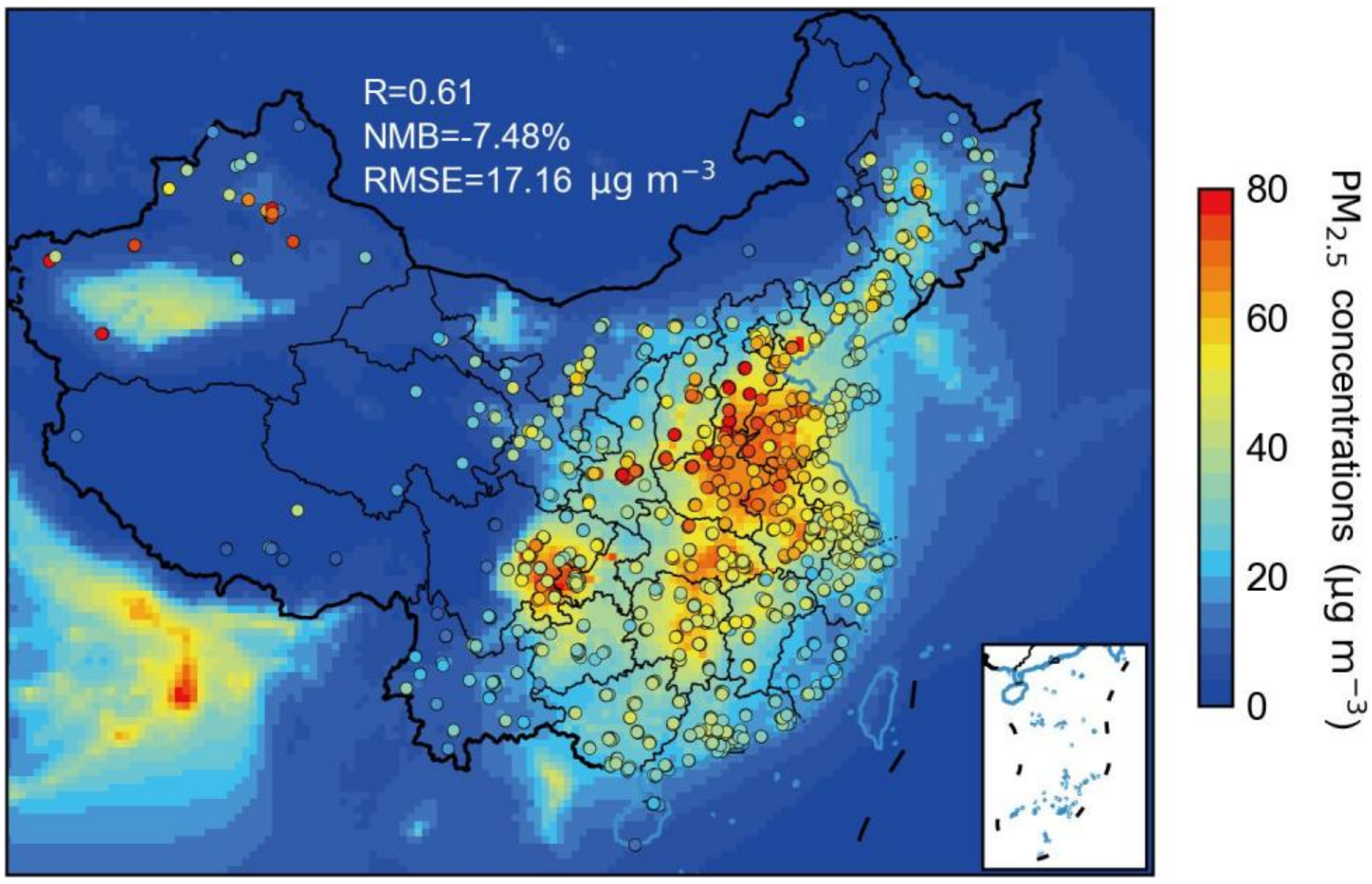

**Supplementary Figure 7 | Evaluation of CMAQ-simulated annual mean $PM_{2.5}$ concentrations in the baseline scenario against surface $PM_{2.5}$ observations.** Measured annual mean $PM_{2.5}$ concentrations in CNEMC stations were overlayed as colored dots. Statistical metrics (R, NMB, and RMSE; see *Methods*) are provided.

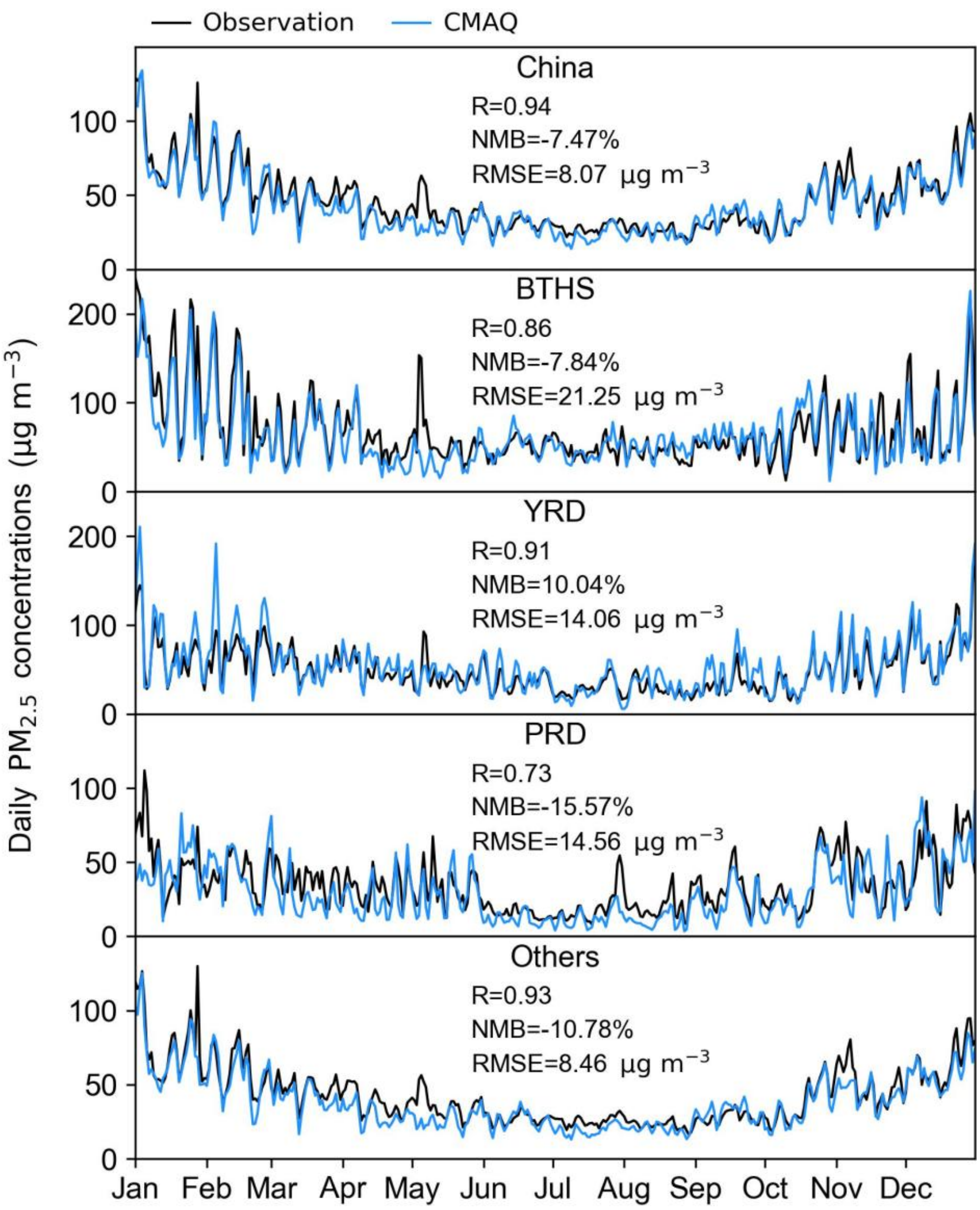

**Supplementary Figure 8 | Evaluation of CMAQ-simulated daily mean $PM_{2.5}$ concentrations against observations in the baseline scenario over China and key regions.** The black lines and blues lines represent concentrations from observation and CMAQ, respectively. Statistical metrics (R, NMB, and RMSE; see *Methods*) are provided.

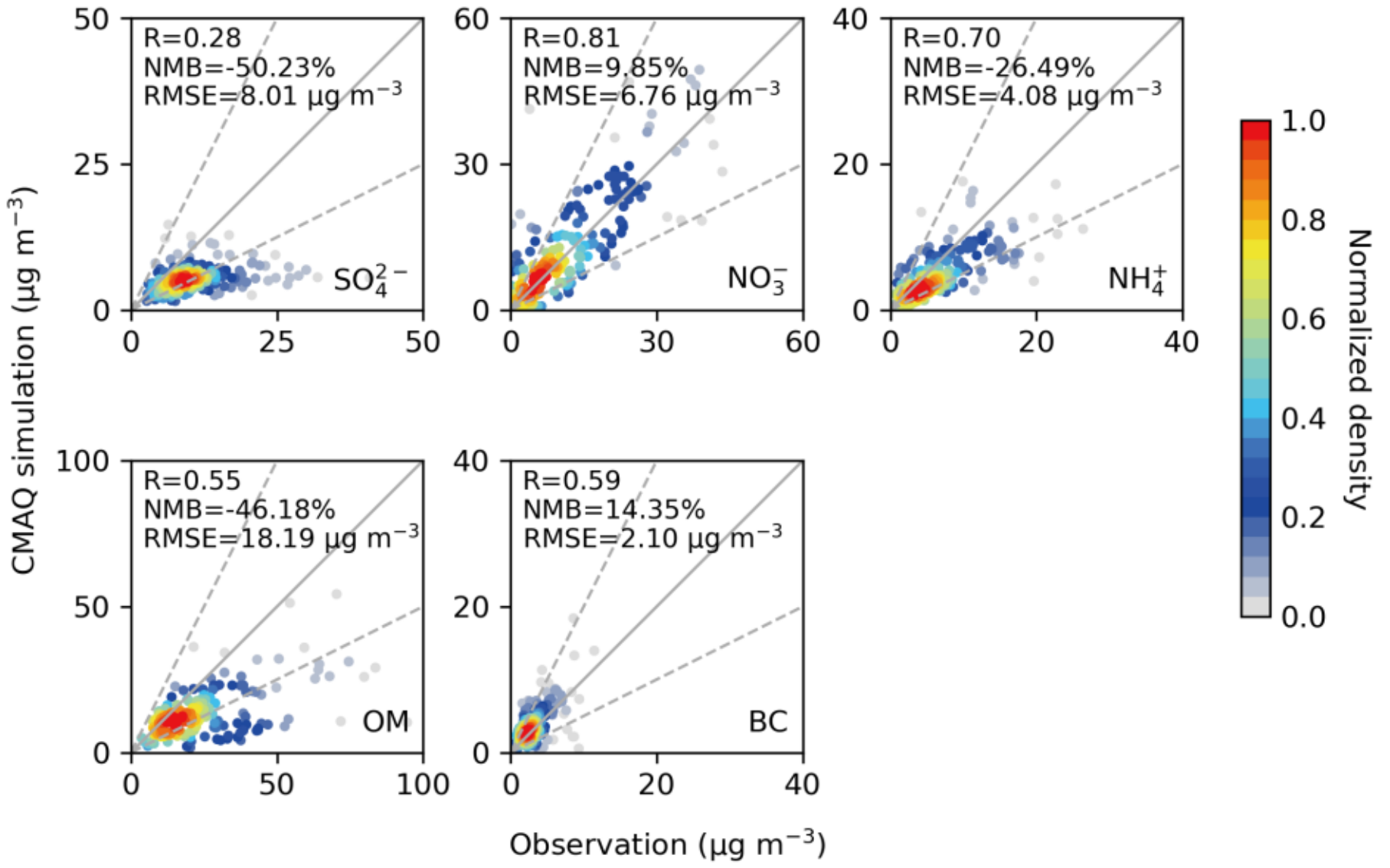

**Supplementary Figure 9 | Evaluation of simulated $PM_{2.5}$ chemical composition concentrations against ground measurements.** The solid line is the 1:1 line, and the dashed lines represent the 1:2 and 2:1 lines. Statistical metrics (R, NMB, and RMSE; see *Methods*) are provided.

**Supplementary Table 1 | Detailed variables included in CleanAir's inputs and outputs.**

| | Features | Categories | Variables |
|---|---|---|---|
| Baseline scenario (inputs) | $PM_{2.5}$ components | 2D variables at 127 × 172 grid | $SO_4^{2-}$, $NO_3^-$, $NH_4^+$, OM, SOM, BC, $PM_{other}$ |
| | Other relevant species | 2D variables at 127 × 172 grid | MDA8 $O_3$, MDA1 $O_3$, Daily mean $O_3$ |
| | | 3D variables at 127 × 172 grid | OH, $HO_2$, $H_2O_2$, $O_3$, $HNO_3$, HONO, $N_2O_5$, $NH_3$ |
| | Meteorological fields | 2D variables at 127 × 172 grid | Surface pressure, planetary boundary layer height, sensible heat flux, latent heat flux, inverse of aerodynamic resistance, skin temperature at ground, temperature at 2 m, mixing ratio at 2 m, wind speed at 10 m, wind speed at 10 m, longwave radiation at ground, solar radiation absorbed at ground, solar rad reaching surface, non-convective precipitation, convective precipitation, total cloud fraction, cloud bottom layer height, cloud top layer height, leaf-area index, sea ice fraction, and soil texture type by USDA category |
| | | 3D variables at 127 × 172 grid | Density of air, air temperature, water vapor mixing ratio, pressure, mid-layer height above ground, 3D resolved cloud fraction |
| | | 3D variables at 128 × 173 grid | U-component of true wind, V-component of true wind |
| | Biogenic emissions | 2D variables at 127 × 172 grid | NMVOCs |
| | Anthropogenic emissions | Elevated layer; 2D variables at 127 × 172 grid | $PM_{2.5}$, BC, OC, $PM_{coarse}$, NMVOCs, $NH_3$, $NO_x$, $SO_2$ |
| | | Intermediate layer; 2D variables at 127 × 172 grid | $PM_{2.5}$, BC, OC, $PM_{coarse}$, NMVOCs, $NH_3$, $NO_x$, $SO_2$ |
| | | Surface layer; 2D variables at 127 × 172 grid | $PM_{2.5}$, BC, OC, $PM_{coarse}$, NMVOCs, $NH_3$, $NO_x$, $SO_2$ |
| | | All emissions outside China; 2D variables at 127 × 172 grid | $PM_{2.5}$, BC, OC, $PM_{coarse}$, NMVOCs, $NH_3$, $NO_x$, $SO_2$ |
| Emission reduction scenario (inputs) | Anthropogenic emissions | Elevated layer; 2D variables at 127 × 172 grid | $PM_{2.5}$, BC, OC, $PM_{coarse}$, NMVOCs, $NH_3$, $NO_x$, $SO_2$ |
| | | Intermediate layer; 2D variables at 127 × 172 grid | $PM_{2.5}$, BC, OC, $PM_{coarse}$, NMVOCs, $NH_3$, $NO_x$, $SO_2$ |
| | | Surface layer; 2D variables at 127 × 172 grid | $PM_{2.5}$, BC, OC, $PM_{coarse}$, NMVOCs, $NH_3$, $NO_x$, $SO_2$ |
| Emission reduction scenario (outputs) | $PM_{2.5}$ components | 2D variables at 127 × 172 grid | $SO_4^{2-}$, $NO_3^-$, $NH_4^+$, OM, SOM, BC, $PM_{other}$ |
| | Other reactants | 2D variables at 127 × 172 grid | MDA8 $O_3$, MDA1 $O_3$, Daily mean $O_3$ |

**Supplementary Table 2 | Anthropogenic emissions of major pollutants in China in 2017 provided by MEIC-HR (unit: Tg).**

| Sector | $PM_{2.5}$ | BC | OC | $PM_{10}$ | $SO_2$ | $NO_x$ | $NH_3$ | NMVOCs |
|---|---|---|---|---|---|---|---|---|
| Total | 7.6 | 1.3 | 2.1 | 10.2 | 10.5 | 22.0 | 10.3 | 28.5 |
| Power | 0.6 | 0.0 | 0.0 | 1.0 | 1.8 | 4.2 | 0.0 | 0.1 |
| Industry | 3.5 | 0.3 | 0.3 | 5.2 | 6.0 | 9.2 | 0.3 | 19.0 |
| Transportation | 0.5 | 0.3 | 0.1 | 0.6 | 0.3 | 7.7 | 0.0 | 4.8 |
| Residential | 3.0 | 0.6 | 1.7 | 3.4 | 2.4 | 0.8 | 0.3 | 4.6 |
| Agriculture | 0.0 | 0.0 | 0.0 | 0.0 | 0.0 | 0.0 | 9.6 | 0.0 |

**Supplementary Table 3 | Definition of key regions.**

| Region name | Abbreviation | Cities |
|---|---|---|
| Beijing-Tianjin-Hebei and its surroundings | BTHS | Beijing; Tianjin; Shijiazhuang, Tangshan, Langfang, Baoding, Cangzhou, Hengshui, Xingtai, and Handan in Hebei province; Taiyuan, Yangquan, Changzhi, and Jincheng in Shanxi province; Jinan, Zibo, Jining, Dezhou, Liaocheng, Binzhou, and Heze in Shandong province; Zhengzhou, Kaifeng, Anyang, Hebi, Xinxiang, Jiaozuo, and Puyang in Henan province. |
| Yangtze River Delta | YRD | Shanghai; Nanjing, Wuxi, Xuzhou, Changzhou, Suzhou, Nantong, Lianyungang, Huaian, Yancheng, Yangzhou, Zhenjiang, Taizhou (泰州), and Suqian in Jiangsu province; Hangzhou, Ningbo, Wenzhou, Shaoxing, Huzhou, Jiaxing, Jinhua, Quzhou, Taizhou (台州), Lishui, and Zhoushan in Zhejiang province; Hefei, Wuhu, Bengbu, Huainan, Maanshan, Huaibei, Tonglin, Anqing, Huangshan, Fuyang, Suzhou, Chuzhou, Luan, Xuancheng, Chizhou, and Bozhou in Anhui province. |
| Pearl River Delta | PRD | Guangzhou, Shenzhen, Zhuhai, Foshan, Jiangmen, Zhaoqing, Huizhou, Dongguan, and Zhongshan in Guangdong province. |
| Region for short-term emission control | / | Beijing; Tianjin; Shijiazhuang, Tangshan, Qinhuangdao, Handan, Xingtai, Baoding, Zhangjiakou, Chengde, Cangzhou, Langfang, Hengshui in Hebei province; Jinan, Qingdao, Zibo, Zaozhuang, Dongying, Yantai, Weifang, Jining, Taian, Weihai, Rizhao, Linyi, Dezhou, Liaocheng, Binzhou, Heze in Shandong province; Taiyuan, Datong, Yangquan, Changzhi, Jincheng, Shuozhou, Jinzhong, Yuncheng, Yizhou, Linfen, Lvliang in Shanxi province; Zhengzhou, Kaifeng, Luoyang, Pingdingshan, Anyang, Hebi, Xinxiang, Jiaozuo, Puyang, Xuchang, Luohe, Sanmenxia, Nanyang, Shangqiu, Xinyang, Zhoukou, Zhumadian in Henan province. |

**Supplementary Table 4 | Evaluation of meteorological variables in the baseline scenario simulated by the WRF model.**

| | Sample Number | R | Mean Obs. | Mean Sim. | MB | RMSE | NMB | NME |
|---|---|---|---|---|---|---|---|---|
| Temperature (°C) | 5368277 | 0.97 | 15.12 | 14.34 | -0.78 | 3.30 | -5.14 | 16.21 |
| Relative humidity (%) | 5359708 | 0.73 | 69.40 | 70.98 | 1.59 | 15.48 | 2.29 | 16.92 |
| Wind speed (m/s) | 5178511 | 0.59 | 2.71 | 3.15 | 0.43 | 2.04 | 15.96 | 56.89 |
| Wind direction (°) | 4340632 | 0.38 | 197.33 | 191.37 | 4.72 | 68.70 | 2.39 | 25.67 |
| Precipitation (mm) | 236566 | 0.39 | 5.79 | 6.22 | 0.43 | 15.16 | 7.38 | 117.52 |

Note: (1) Units for Mean Observation (Mean Obs.), Mean Simulation (Mean Sim.), Mean Bias (MB), and Root Mean Square Error (RMSE) are shown below the name of each variable in the first column. Units for Normalized Mean Bias (NMB) and Normalized Mean Error (NME) are %. (2) Mean bias in wind direction is calculated considering the periodic nature of wind direction (e.g., the difference between 1° and 359° is treated as 2°, rather than 358°).

**Supplementary Table 5 | Evaluation of $PM_{2.5}$ concentrations in the baseline scenario simulated by the CMAQ model.**

| | Sample Number | R | Mean Obs. | Mean Sim. | MB | RMSE | NMB | NME |
|---|---|---|---|---|---|---|---|---|
| China | 528534 | 0.64 | 45.68 | 42.27 | -3.41 | 32.84 | -7.46 | 44.73 |
| BTHS | 52500 | 0.68 | 68.37 | 63.21 | -5.16 | 38.45 | -7.55 | 36.37 |
| YRD | 70866 | 0.78 | 48.60 | 53.49 | 4.88 | 25.45 | 10.04 | 36.21 |
| PRD | 19664 | 0.65 | 34.27 | 28.94 | -5.33 | 18.75 | -15.56 | 39.98 |
| Others | 385504 | 0.59 | 42.64 | 38.04 | -4.59 | 33.77 | -10.77 | 48.54 |

Note: Unit for Mean Observation (Mean Obs.), Mean Simulation (Mean Sim.), Mean Bias (MB), Root Mean Squared Error (RMSE), Normalized Mean Bias (NMB), and Normalized Mean Error (NME) are $\mu g\ m^{-3}$, $\mu g\ m^{-3}$, $\mu g\ m^{-3}$, $\mu g\ m^{-3}$, %, and %, respectively.